\documentstyle[12pt,epsf]{article}
\thicklines
\begin{document}

\vspace*{3cm}
\begin{center}
{\bf\sf  THEORETICAL STATUS OF THE $\sf B_c$ MESON}\\

\vspace*{1cm}
{\sc S.S.Gershtein, \underline{V.V. Kiselev}\footnote{
A speaker of the talk given at IV Workshop on Heavy Quark Physics,
Rostock, September 19-22, 1997.}, A.K.Likhoded, A.V.Tkabladze,\\
A.V.Berezhnoy and A.I.Onishchenko}\\

\vspace*{0.7cm}
IHEP, Protvino, Russia\\

~~~
\end{center}
\begin{abstract}
Theoretical predictions on the $B_c$ meson properties are reviewed.
\end{abstract}

\vspace*{3cm}
1. Introduction

2. Mass Spectrum of the $(\bar b c)$ family

3. Leptonic constants

4. $B_c$ decays

5. $B_c$ production

6. Conclusions

\newpage
\centerline{\bf\sc 1. Introduction}

\vspace*{1cm}
Theoretical and experimental studies of  the heavy quark sector
in the Standard Model are of a great interest to complete the whole
quantitative picture of fundamental interactions.

In the bottom quark physics experimentalists step from the
\underline{$10^6$} yield of hadrons containing $b$-quark at present
facilities up to the \underline{$10^9$} yield in the foreseeable future
to measure rare processes
like the {\sf CP}-violation and possible effects beyond the SM. To distinguish
the hadronic dynamics from the latter effects at the quark level one needs
the perfect understanding of QCD interactions binding the quarks into hadrons.

One of the accompanying problem is the observation and study of the $(\bar b
c)$
state, yielding a $10^{-3}$ fraction of beauty hadrons at high energies.

The $B_c$ meson allows one

1. to accomplish the QCD-based models of hadrons with the bottom quarks,

2. to study the specific production and decay mechanisms,

3. to measure the SM parameters.

\vspace*{5mm}
The basic state of $B_c$ is the long-lived heavy quarkonium, which can be
searched for in a way analogous to the observation of beauty mesons
with a light quark \cite{ufn}.

\noindent
At CDF the background is still strong to isolate the $B_c$ event at low
statistics available \cite{cdf}.
At present, the LEP Collaborations have reported on several candidates for the
$B_c$ decays \cite{lep}.

\begin{table}[h]
\caption{The basic characteristics of events with the $B_c$-meson candidates at
LEP.}
\begin{center}
\begin{tabular}{||l|c|c|c||}
\hline
Collab.& Mode & mass of $B_c$, GeV & lifetime, ps\\
\hline
OPAL$_{old}$  & $\psi\pi^{\pm}$  & $6.31\pm 0.17$ & --- \\
OPAL$_{new}$  & $\psi\pi^{\pm}$  & $6.29\pm 0.17$ & $-0.06\pm 0.19$ \\
OPAL$_{new}$  & $\psi\pi^{\pm}$  & $6.33\pm 0.06$ & $0.09\pm 0.10$ \\
ALEPH & $\psi l^{\pm}\nu$& $5.96\pm 0.24$ & $1.77\pm 0.17$ \\
DELPHI& $\psi\pi^{\pm}$  & $6.35\pm0.09 $ & $0.38\pm 0.06$ \\
DELPHI& $\psi(3\pi)^{\pm}$&$6.12\pm0.02 $ & $0.41\pm 0.07$ \\
\hline
\end{tabular}
\end{center}
\end{table}
The mean values averaged over the $\psi\pi$ mode are equal to
\begin{eqnarray}
m_{B_c} & = & 6.33\pm 0.05\;\; {\rm GeV,}\\
\tau_{B_c} & = & 0.28^{+0.10}_{-0.20}\;\; {\rm ps.}
\end{eqnarray}

OPAL has reported also 
$$
f(\bar b \to B_c^+)\cdot {\rm BR}(B_c^+\to \psi \pi^+) =
(3.8^{+5.0}_{-2.4}\pm 0.5)\cdot 10^{-5}.
$$

The question of the report is

\vspace*{3mm}
\centerline{\sf What are the theoretical
expectations of the $\sf B_c$-meson properties?}

\newpage

\centerline{\bf\sc 2. The mass spectrum of the $\bf (\bar b c)$ family}

\vspace*{4mm}
The most accurate estimates of $(\bar b c)$ masses \cite{ger,eq} can be
obtained in the framework of nonrelativistic potential models based on the
NRQCD expansion over both $1/m_Q$ and $v_{rel}\to 0$ \cite{bra}.

The uncertainty of evaluation is about 30 MeV. The reason is
the following. The potential models \cite{pot} were justified for the well
measured masses of charmonium and bottomonium. So, the potentials with
various global behaviour, i.e. with the different $r\to \infty$ and
$r\to 0$ asymptotics, have the same form in the range of mean distances
between the quarks in the heavy quarkonia at $0.2 < r < 1$ fm \cite{eic}.
The observed {\sf regularity} is the distances between the excitation levels
are approximately flavor-independent. The latter is exact for the
logarithmic potential (the Feynman--Hell-Mann theorem), where the
average kinetic energy of quarks $T$ is a constant value independent of
the excitation numbers (the virial theorem) \cite{log}. 
A slow dependence of the
level distances on the reduced mass can be taken into account by the use of
the \underline{Martin} potential (power law:
$V(r)=A(r/r_0)^a+C$, $a\ll 1$) \cite{mart}, wherein the predictions are
in agreement with the QCD-motivated
\underline{Buchm\" uller-Tye} potential  with the account for the two-loop 
evolution of the coupling constant at short distances \cite{bt}.

So, one gets the picture of $(\bar b c)$ levels which is very close to the
texture of charmonium and bottomonium.
The difference is the {\sf jj}-binding instead of the
{\sf LS} one.

The spin-dependent perturbation of the potential includes the
contribution of the effective one-gluon exchange (the vector part)
as well as the scalar confining term \cite{fein}.
\begin{eqnarray}
   V_{SD}(\vec{r}) & = &\biggl(\frac{\vec{L}\cdot\vec{S}_c}{2m_c^2} +
\frac{\vec{L}\cdot\vec{S}_b}{2m_b^2}\biggr)\;
\biggl(-\frac{dV(r)}{rdr}+\frac{8}{3}\;\alpha_s\;\frac{1}{r^3}\biggr) 
+ \nonumber \\
~ & ~ & +\frac{4}{3}\;\alpha_s\;\frac{1}{m_c
m_b}\;\frac{\vec{L}\cdot\vec{S}}{r^3}
+\frac{4}{3}\;\alpha_s\;\frac{2}{3m_c m_b}\;
\vec{S}_c\cdot\vec{S}_b\;4\pi\;\delta(\vec{r}) \label{3} \\
~ & ~ & +\frac{4}{3}\;\alpha_s\;\frac{1}{m_c m_b}\;(3(\vec{S}_c\cdot\vec{n})\;
(\vec{S}_b\cdot\vec{n}) - \vec{S}_c\cdot\vec{S}_b)\;\frac{1}{r^3}\;,
\;\;\vec{n}=\frac{\vec{r}}{r}\;.\nonumber
\end{eqnarray}

The model-dependent value of effective $\alpha_s$ \cite{eq} can
be extracted from the data on the splitting in the charmonium
$$
M(\psi)-M(\eta_c) = \alpha_s \frac{8}{9m_c^2} |R(0)|^2\approx 117 \;
{\rm MeV.}
$$
We take into account the renormalization-group dependence
of $\alpha_s$ at the one-loop accuracy by means of introduction of
the quarkonium scale \cite{ger}
$$
\mu^2 = \langle {\bf p}^2\rangle =2 \langle T\rangle m_{red}.
$$
The estimated difference between the masses of basic pseudoscalar state
and its vector excitation \cite{ger} is equal to 
$$
M(B_c^{*+})-M(B_c^+)=65\pm 15\; {\rm MeV.}
$$
The mass of the ground state \cite{ger} equals 
\begin{equation}
M(B_c^+)=6.25\pm 0.03\; {\rm GeV,}
\end{equation}
which is inside the quoted region of the $B_c$ candidates at LEP.

\begin{figure}[p]
\begin{center}
\begin{picture}(175,150)
\put(15,30){\line(1,0){20}}
\put(15,33){$1S$}
\put(36,25){\line(1,0){20}}
\put(36,32){\line(1,0){20}}

\put(57,21){$^{0^-}$}
\put(57,29){$^{1^-}$}

\put(15,89){\line(1,0){20}}
\put(15,92){$2S$}
\put(36,87){\line(1,0){20}}
\put(36,90){\line(1,0){20}}

\put(57,83){$^{0^-}$}
\put(57,87){$^{1^-}$}

\put(15,124){\line(1,0){20}}
\put(15,127){$3S$}

\put(70,73){\line(1,0){20}}
\put(70,76){$2P$}
\put(91,68){\line(1,0){20}}
\put(91,72){\line(1,0){20}}
\put(91,74){\line(1,0){20}}
\put(91,76){\line(1,0){20}}

\put(112,63){$^{0^+}$}
\put(112,75){$^{2^+}$}
\put(112,67){$^{1^+}$}
\put(112,71){$^{1'^+}$}

\put(70,112){\line(1,0){20}}
\put(70,115){$3P$}
\put(91,109){\line(1,0){20}}
\put(91,111){\line(1,0){20}}
\put(91,112){\line(1,0){20}}
\put(91,114){\line(1,0){20}}

\put(112,102){$^{0^+}$}
\put(112,114){$^{2^+}$}
\put(112,106){$^{1^+}$}
\put(112,110){$^{1'^+}$}

\put(70,140){\line(1,0){20}}
\put(70,143){$4P$}

\put(125,101){\line(1,0){20}}
\put(125,104){$3D$}
\put(146,099){\line(1,0){20}}
\put(146,100){\line(1,0){20}}
\put(146,101){\line(1,0){20}}
\put(146,102){\line(1,0){20}}

\put(167,099){$^{1^-}$}
\put(167,095){$^{3^-}$}
\put(167,091){$^{2^-}$}
\put(167,103){$^{2'^-}$}

\put(125,131){\line(1,0){20}}
\put(125,134){$4D$}

\put(10,0){\framebox(165,150)}
\put(0,0){$6.0$}
\put(10,50){\line(1,0){3}}
\put(0,50){$6.5$}
\put(10,100){\line(1,0){3}}
\put(0,100){$7.0$}
\put(0,150){$7.5$}
\put(0,160){GeV}
\put(10,115){\line(1,0){55}}
\put(10,115.3){\line(1,0){55}}
\put(120,115){\line(1,0){55}}
\put(120,115.3){\line(1,0){55}}
\put(130,117){$BD$ threshold}
\end{picture}
\end{center}
\caption{The mass spectrum of $(\bar b c)$ with account for the
spin-dependent splittings.}
\end{figure}
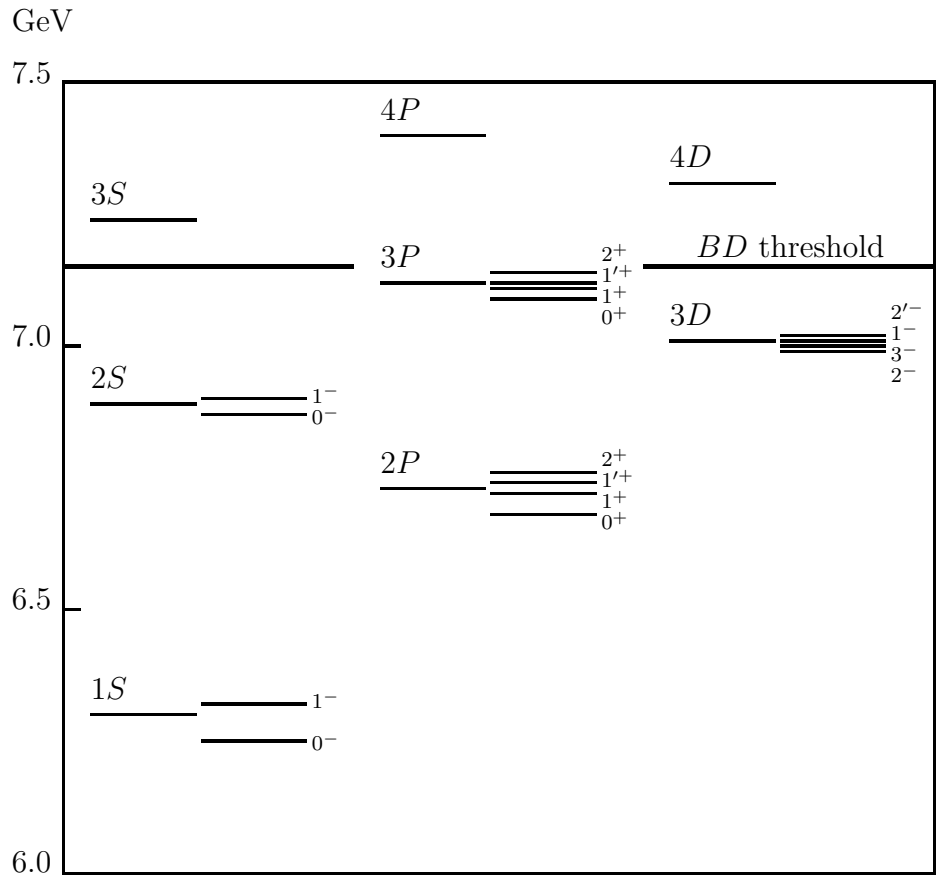

\begin{table}[p]
\caption{The masses of bound $(\bar b c)$-states below the threshold of decay
into the pair of heavy mesons $BD$, in GeV, in the models with the Martin 
and BT potentials. The spectroscopic 
notations of states are $n^{2j_c}L_J$, where $j_c$ is the total angular 
momentum of $c$-quark, $n$ is the principal quantum number, $L$ is the 
orbital angular momentum, $J$ is the total spin of meson.}
\label{t1}
\begin{center}
\begin{tabular}{||l|c|c||}
\hline
state & Martin  & BT \\
\hline
$1^1S_0$    & 6.253       & 6.264    \\
$1^1S_1$    & 6.317       & 6.337    \\
$2^1S_0$    & 6.867       & 6.856    \\
$2^1S_1$    & 6.902       & 6.899    \\
$2^1P_0$    & 6.683       & 6.700    \\
$2P\; 1^+$  & 6.717       & 6.730    \\
$2P\; 1'^+$ & 6.729       & 6.736    \\
$2^3P_2$    & 6.743       & 6.747    \\
$3^1P_0$    & 7.088       & 7.108    \\
$3P\; 1^+$  & 7.113       & 7.135    \\
$3P\; 1'^+$ & 7.124       & 7.142    \\
$3^3P_2$    & 7.134       & 7.153    \\
$3D\; 2^-$  & 7.001       & 7.009    \\
$3^5D_3$    & 7.007       & 7.005    \\
$3^3D_1$    & 7.008       & 7.012    \\
$3D\; 2'^-$ & 7.016       & 7.012    \\
\hline
\end{tabular}
\end{center}
\end{table}

\centerline{\bf\sc Radiative transitions}

\vspace*{3mm}
{\sf The bright feature of the $(\bar b c)$ family is that there are no
annihilation
decay modes due to the strong interaction.} So, the excitations, in a cascade
way,
decay into the ground state with the emission of {\sf photons} and {\sf
pion-pion} pairs.

The formulae for the {\sf E1}-transitions are slightly modified.
\begin{eqnarray}
\Gamma(\bar nP_J\to n^1S_1 +\gamma) & = & \frac{4}{9}\;\alpha_{{\rm
em}}\;Q^2_{{\rm eff}}\;
       \omega^3\;I^2(\bar nP;nS)\;w_J(\bar nP) \;,\nonumber\\
\Gamma(\bar nP_J\to n^1S_0 +\gamma) & = & \frac{4}{9}\;\alpha_{{\rm
em}}\;Q^2_{{\rm eff}}\;
       \omega^3\;I^2(\bar nP;nS)\;(1-w_J(\bar n P)) \;,\nonumber\\
\Gamma(n^1S_1\to \bar n P_J +\gamma) & = & \frac{4}{27}\;\alpha_{{\rm
em}}\;Q^2_{{\rm eff}}\;
       \omega^3\;I^2(nS;\bar n P)\;(2J+1)\;w_J(\bar n P)\;, \label{7} \\
\Gamma(n^1S_0\to \bar n P_J +\gamma) & = & \frac{4}{9}\;\alpha_{{\rm
em}}\;Q^2_{{\rm eff}}\;
       \omega^3\;I^2(nS;\bar n P)\;(2J+1)\;(1-w_J(\bar n P))\;,\nonumber\\
\Gamma(\bar n P_J\to n D_{J'} +\gamma) & = & \frac{4}{27}\;\alpha_{{\rm
em}}\;Q^2_{{\rm eff}}\;
       \omega^3\;I^2(nD;\bar n P)\;(2J'+1)\;
\nonumber\\ && w_J(\bar n P)) w_{J'}(nD)
        S_{JJ'}\;,\nonumber\\
\Gamma(n D_J\to \bar n P_{J'} +\gamma) & = & \frac{4}{27}\;\alpha_{{\rm
em}}\;Q^2_{{\rm eff}}\;
       \omega^3\;I^2(nD;\bar n P)\;(2J'+1)\;
\nonumber\\ && w_{J'}(\bar n P)) w_{J}(nD)
        S_{J'J}\;,\nonumber
\end{eqnarray}
where $\omega$ is the photon energy, $\alpha_{\rm em}$ is the electromagnetic
fine structure constant. 
In eq.(\ref{7}) one uses
\begin{equation}
   Q_{{\rm eff}}=\frac{m_c Q_{\bar b} - m_b Q_c }{m_c +m_b}\;, 
\end{equation}
where $Q_{c,b}$ are the electric charges of the quarks. For the
$B_c$ meson with the parameters from the Martin potential, one gets 
$Q_{\rm eff}=0.41$. 
$w_J(nL)$ is the probability that the spin $S=1$ 
in the $nL$ state. $S_{JJ'}$ are the statistical factors.
The $I(\bar nL;nL')$ value is expressed through the radial wave functions,
\begin{equation}
   I(\bar n L;nL') = |\int R_{\bar n L}(r) R_{nL'}(r) r^3 {\rm d}r|\;. 
\label{7aa}
\end{equation}

For the dipole magnetic {\sf M1}-transitions one has 
\begin{equation}
\Gamma(\bar n^1S_i\to n^1S_f +\gamma) =  \frac{16}{3}\;\mu^2_{{\rm
eff}}\;\omega^3\;
(2f+1)\;A_{if}^2\;,
\label{10}
\end{equation}
where 
$$ A_{if} = \int R_{\bar n S}(r) R_{nS}(r) j_0(\omega r/2) r^2\; {\rm d}r\;,
$$
and
\begin{equation}
\mu_{{\rm eff}}=\frac{1}{2}\;\frac{\sqrt{\alpha_{{\rm em}}}}{2m_c m_b}\;
(Q_c m_b - Q_{\bar b} m_c)\;.
\label{11}
\end{equation}
Note, in contrast to the $\psi$  and $\Upsilon$ particles, the total width of
the $B_c^*$ meson is equal to the width of its radiative decay into
the $B_c(0^-)$ state.

Thus, below the threshold of decay into the BD-pair
the theory predicts the existence of 16
narrow $(\bar b c)$ states, which do not annihilate due to the strong
interactions, but they have the cascade radiative transitions into the ground
long-lived pseudoscalar state, the $B_c^+$ meson.
\newpage
\begin{table}[ph]
\caption{The total widths of excited bound $(\bar b c)$-states below the
threshold of decay into the $BD$-pair in the model with Martin potential
and the branching ratios of the dominant decay modes.}
\label{t2}
\begin{center}
\begin{tabular}{||l|r|p{29.3mm}|r||}
\hline
state & $\Gamma_{\rm tot}$, KeV  &dominant decay mode & BR, \% \\
\hline
$1^1S_1$    & 0.06       & $1^1S_0+\gamma$   & 100  \\
$2^1S_0$    & 67.8       & $1^1S_0+\pi\pi$   &  74  \\
$2^1S_1$    & 86.3       & $1^1S_1+\pi\pi$   &  58  \\
$2^1P_0$    & 65.3       & $1^1S_1+\gamma$   & 100  \\
$2P\; 1^+$  & 89.4       & $1^1S_1+\gamma$   &  87  \\
$2P\; 1'^+$ & 139.2      & $1^1S_0+\gamma$   &  94  \\
$2^3P_2$    & 102.9      & $1^1S_1+\gamma$   & 100  \\
$3^1P_0$    & 44.8       & $2^1S_1+\gamma$   &  57  \\
$3P\; 1^+$  & 65.3       & $2^1S_1+\gamma$   &  49  \\
$3P\; 1'^+$ & 92.8       & $2^1S_0+\gamma$   &  63  \\
$3^3P_2$    & 71.6       & $2^1S_1+\gamma$   &  69  \\
$3D\; 2^-$  & 95.0       & $2P\; 1^++\gamma$ &  47  \\
$3^5D_3$    & 107.9      & $2^3P_2+\gamma$   &  71  \\
$3^3D_1$    & 155.4      & $2^1P_0+\gamma$   &  51  \\
$3D\; 2'^-$ & 122.0      & $2P\; 1'^++\gamma$ & 38 \\
\hline
\end{tabular}
\end{center}
\end{table}
As for the states lying above the threshold of decay into the heavy meson 
$BD$ pair, the width of $B_c^{*+}(3S)\to B^+D^0$, for example, can be
calculated in the framework of sum rules for the meson currents, where the
scaling relation \cite{g-sr} takes place for the $g$ constants of similar
decays of
$\Upsilon(4S)\to B^+B^-$ and $\psi(3770)\to D^+D^-$,
$$
\frac{g^2}{M}\; \biggl(\frac{4m_{12}}{M}\biggl)= const.
$$
The relation is caused by the dependence of energy gap between the vector and 
pseudoscalar heavy meson states: $\Delta M_{1,2}\cdot M_{1,2} = const.$,
where $M_{1,2}$ are the meson masses in the final state, $m_{12}$ is
their reduced mass. The width of this decay has a strong dependence
on the $B_c^{*+}(3S)$ mass, and at $M=7.25$ GeV it is equal to 
$\Gamma= 90\pm 35$ MeV, where the uncertainty is determined by the accuracy of 
method used.
\vspace*{5mm}

\begin{table}[bh]
\caption{The predictions of scaling relation in comparison with the 
current experimental data}
\begin{center}
\begin{tabular}{||l|c|c||}
\hline
value & exp. & scaling rel.\\
\hline
$g_{\Upsilon(4S)\to B^+B^-}$ & 52 & input\\
$g_{\psi(3770)\to D^+D^-}$   & 31 & 31\\
$g_{B_c^{*+}(3S)\to B^+D^0}$ & -- & 49\\
\hline
\end{tabular}
\end{center}
\end{table}

\centerline{\bf\sc 3. Leptonic constants}

\vspace*{5mm}

In the framework of \underline{potential models} the
asymptotic behaviour at $r\to \infty$, $r\to 0$ are significant for the 
determination of the leptonic 
coupling constants $f$ for the $nS$-levels. In the leading 
approximation, the $f$ value does not depend on the spin state of the level 
and it is determined by the value of the radial wave function at the origin, 
$R(0)$, being model-dependent,
$$
\tilde f_{n} = \sqrt{\frac{3}{\pi M_{n}}} R_{nS}(0)\;.
$$
Taking into account the hard gluon corrections, the constants 
of vector and pseudoscalar states equal
\begin{equation}
f_{n}^{V,P} = \tilde f_{n}\; \biggl(1+\frac{\alpha_s}{\pi}
\biggl(\frac{m_1-m_2}{m_1+m_2}\ln\frac{m_1}{m_2}-\delta^{V,P}\biggr)\biggr)\;,
\end{equation}
where $m_{1,2}$ are the quark masses, $\delta^V=8/3$, $\delta^P=2$
\cite{f0,scale,vol},  and
the QCD coupling constant is determined at the scale of the quark masses.

The corresponding uncertainty due to the model dependence is expressed
by a factor of two.

\begin{table}[th]
\caption{The radial wave functions at the origin, $R_{nS}(0)$ (in GeV$^{3/2}$)
and $R'_{nP}(0)$ (in GeV$^{5/2}$), obtained in the Schr\" odinger
equation with the Martin and BT potentials as well as in the sum rules.}
\label{t3}
\begin{center}
\begin{tabular}{||c|c|c|c||}
\hline
n & Martin  & BT & SR\\
\hline
$R_{1S}(0)$  & 1.31 & 1.28 & 1.20\\
$R_{2S}(0)$  & 0.97 & 0.99 & 0.85\\
$R'_{2P}(0)$ & 0.55 & 0.45 & --\\
$R'_{3P}(0)$ & 0.57 & 0.51 & --\\
\hline
\end{tabular}
\end{center}
\end{table}

The \underline{QCD sum rules} \cite{sr} allow one to determine the leptonic
constants for the heavy quarkonium states with a much better accuracy.

Standard schemes of the sum 
rules give an opportunity to calculate the ground state constants for 
vector and pseudoscalar currents with the account for corrections
from the quark-gluon condensates, which have the power form over the inverse
heavy quark mass.

\vspace*{2mm}
\centerline{\bf\sc Quasilocal sum rules}

\vspace*{2mm}
There is a region of the momentum numbers
for the spectral density of the two-point current correlator, where the
condensate contributions are not significant. In this region,
the \underline{integral representation} for the contribution
of resonances lying bellow the threshold of decay into the pair of
heavy mesons, allows one to use the {\sf regularity} of the quarkonium state
density mentioned above, and to derive the \underline{scaling relations}
\cite{scale} for the leptonic  constants of the ground state quarkonia with
different quark contents and for the excited states. 
Thus, for vector states we have
\begin{equation}
\frac{f^2_n}{M_n}\;\biggl(\frac{M_n}{M_1}\biggr)^2\;
\biggl(\frac{m_1+m_2}{4m_{12}}\biggr)^2 = \frac{c}{n}\;,
\label{scal}
\end{equation}
where $m_{12}=m_1m_2/(m_1+m_2)$ is the reduced mass of quarks,
and the constant $c$ is determined by the average kinetic energy of quarks,
the QCD coupling constant at the scale of average momentum transfers
in the system and the hard gluon correction factor to the vector current.
Numerically, in the method accuracy the $c$ value turns out to depend on
no quark flavour and excitation number in the system. Relation (\ref{scal})
is in a good agreement with  the data on the coupling constants for
the families of $\psi$ and $\Upsilon$ particles and, thus, it can be
a reliable basis for the prediction of leptonic constants for the
$B_c$-meson family. The constants for the pseudoscalar states of $nS$-levels
are
determined by the relation
$$
f_n^P = f_n \biggl(1+\frac{2\alpha_s}{3\pi}\biggr)\; \frac{m_1+m_2}{M_n}\;.
$$
\begin{figure}[th]
\begin{center}
\begin{picture}(100,95)
\put(15,10){\framebox(83,80)}
\put(15,30){\line(1,0){3}}
\put(3,30){$200$}
\put(15,50){\line(1,0){3}}
\put(3,50){$400$}
\put(15,70){\line(1,0){3}}
\put(3,70){$600$}
\put(15,20){\line(1,0){3}}
\put(15,40){\line(1,0){3}}
\put(15,60){\line(1,0){3}}
\put(15,80){\line(1,0){3}}

\put(0,93){$f_n$, {\rm MeV}}

\put(25,10){\line(0,1){3}}
\put(35,10){\line(0,1){3}}
\put(45,10){\line(0,1){3}}
\put(55,10){\line(0,1){3}}
\put(65,10){\line(0,1){3}}
\put(75,10){\line(0,1){3}}
\put(85,10){\line(0,1){3}}
\put(25,2){$1$}
\put(35,2){$2$}
\put(45,2){$3$}
\put(55,2){$4$}
\put(65,2){$5$}
\put(75,2){$6$}
\put(85,2){$7$}
\put(95,2){$n$}


\put(25,79.0){\circle*{1.6}}
\put(35,57.1){\circle*{1.6}}
\put(45,51.5){\circle*{1.6}}
\put(55,41.0){\circle*{1.6}}
\put(65,45.6){\circle*{1.6}}
\put(75,34.2){\circle*{1.6}}
\put(55,38.0){\line(0,1){6}}
\put(65,41.0){\line(0,1){9.2}}
\put(75,31.2){\line(0,1){6}}

 \put( 25.10, 78.52){\circle*{0.5}}
 \put( 25.20, 78.15){\circle*{0.5}}
 \put( 25.30, 77.79){\circle*{0.5}}
 \put( 25.40, 77.44){\circle*{0.5}}
 \put( 25.50, 77.09){\circle*{0.5}}
 \put( 25.60, 76.75){\circle*{0.5}}
 \put( 25.70, 76.41){\circle*{0.5}}
 \put( 25.80, 76.07){\circle*{0.5}}
 \put( 25.90, 75.74){\circle*{0.5}}
 \put( 26.00, 75.42){\circle*{0.5}}
 \put( 26.10, 75.10){\circle*{0.5}}
 \put( 26.20, 74.78){\circle*{0.5}}
 \put( 26.30, 74.47){\circle*{0.5}}
 \put( 26.40, 74.16){\circle*{0.5}}
 \put( 26.50, 73.86){\circle*{0.5}}
 \put( 26.60, 73.56){\circle*{0.5}}
 \put( 26.70, 73.27){\circle*{0.5}}
 \put( 26.80, 72.97){\circle*{0.5}}
 \put( 26.90, 72.69){\circle*{0.5}}
 \put( 27.00, 72.40){\circle*{0.5}}
 \put( 27.10, 72.12){\circle*{0.5}}
 \put( 27.20, 71.85){\circle*{0.5}}
 \put( 27.30, 71.57){\circle*{0.5}}
 \put( 27.40, 71.30){\circle*{0.5}}
 \put( 27.50, 71.04){\circle*{0.5}}
 \put( 27.60, 70.77){\circle*{0.5}}
 \put( 27.70, 70.51){\circle*{0.5}}
 \put( 27.80, 70.26){\circle*{0.5}}
 \put( 27.90, 70.00){\circle*{0.5}}
 \put( 28.00, 69.75){\circle*{0.5}}
 \put( 28.10, 69.51){\circle*{0.5}}
 \put( 28.20, 69.26){\circle*{0.5}}
 \put( 28.30, 69.02){\circle*{0.5}}
 \put( 28.40, 68.78){\circle*{0.5}}
 \put( 28.50, 68.54){\circle*{0.5}}
 \put( 28.60, 68.31){\circle*{0.5}}
 \put( 28.70, 68.08){\circle*{0.5}}
 \put( 28.80, 67.85){\circle*{0.5}}
 \put( 28.90, 67.63){\circle*{0.5}}
 \put( 29.00, 67.40){\circle*{0.5}}
 \put( 29.10, 67.18){\circle*{0.5}}
 \put( 29.20, 66.96){\circle*{0.5}}
 \put( 29.30, 66.75){\circle*{0.5}}
 \put( 29.40, 66.53){\circle*{0.5}}
 \put( 29.50, 66.32){\circle*{0.5}}
 \put( 29.60, 66.11){\circle*{0.5}}
 \put( 29.70, 65.91){\circle*{0.5}}
 \put( 29.80, 65.70){\circle*{0.5}}
 \put( 29.90, 65.50){\circle*{0.5}}
 \put( 30.00, 65.30){\circle*{0.5}}
 \put( 30.10, 65.10){\circle*{0.5}}
 \put( 30.20, 64.90){\circle*{0.5}}
 \put( 30.30, 64.71){\circle*{0.5}}
 \put( 30.40, 64.52){\circle*{0.5}}
 \put( 30.50, 64.32){\circle*{0.5}}
 \put( 30.60, 64.14){\circle*{0.5}}
 \put( 30.70, 63.95){\circle*{0.5}}
 \put( 30.80, 63.76){\circle*{0.5}}
 \put( 30.90, 63.58){\circle*{0.5}}
 \put( 31.00, 63.40){\circle*{0.5}}
 \put( 31.10, 63.22){\circle*{0.5}}
 \put( 31.20, 63.04){\circle*{0.5}}
 \put( 31.30, 62.87){\circle*{0.5}}
 \put( 31.40, 62.69){\circle*{0.5}}
 \put( 31.50, 62.52){\circle*{0.5}}
 \put( 31.60, 62.35){\circle*{0.5}}
 \put( 31.70, 62.18){\circle*{0.5}}
 \put( 31.80, 62.01){\circle*{0.5}}
 \put( 31.90, 61.84){\circle*{0.5}}
 \put( 32.00, 61.68){\circle*{0.5}}
 \put( 32.10, 61.51){\circle*{0.5}}
 \put( 32.20, 61.35){\circle*{0.5}}
 \put( 32.30, 61.19){\circle*{0.5}}
 \put( 32.40, 61.03){\circle*{0.5}}
 \put( 32.50, 60.87){\circle*{0.5}}
 \put( 32.60, 60.72){\circle*{0.5}}
 \put( 32.70, 60.56){\circle*{0.5}}
 \put( 32.80, 60.41){\circle*{0.5}}
 \put( 32.90, 60.25){\circle*{0.5}}
 \put( 33.00, 60.10){\circle*{0.5}}
 \put( 33.10, 59.95){\circle*{0.5}}
 \put( 33.20, 59.80){\circle*{0.5}}
 \put( 33.30, 59.66){\circle*{0.5}}
 \put( 33.40, 59.51){\circle*{0.5}}
 \put( 33.50, 59.37){\circle*{0.5}}
 \put( 33.60, 59.22){\circle*{0.5}}
 \put( 33.70, 59.08){\circle*{0.5}}
 \put( 33.80, 58.94){\circle*{0.5}}
 \put( 33.90, 58.80){\circle*{0.5}}
 \put( 34.00, 58.66){\circle*{0.5}}
 \put( 34.10, 58.52){\circle*{0.5}}
 \put( 34.20, 58.39){\circle*{0.5}}
 \put( 34.30, 58.25){\circle*{0.5}}
 \put( 34.40, 58.12){\circle*{0.5}}
 \put( 34.50, 57.98){\circle*{0.5}}
 \put( 34.60, 57.85){\circle*{0.5}}
 \put( 34.70, 57.72){\circle*{0.5}}
 \put( 34.80, 57.59){\circle*{0.5}}
 \put( 34.90, 57.46){\circle*{0.5}}
 \put( 35.00, 57.33){\circle*{0.5}}
 \put( 35.10, 57.20){\circle*{0.5}}
 \put( 35.20, 57.08){\circle*{0.5}}
 \put( 35.30, 56.95){\circle*{0.5}}
 \put( 35.40, 56.83){\circle*{0.5}}
 \put( 35.50, 56.70){\circle*{0.5}}
 \put( 35.60, 56.58){\circle*{0.5}}
 \put( 35.70, 56.46){\circle*{0.5}}
 \put( 35.80, 56.34){\circle*{0.5}}
 \put( 35.90, 56.22){\circle*{0.5}}
 \put( 36.00, 56.10){\circle*{0.5}}
 \put( 36.10, 55.98){\circle*{0.5}}
 \put( 36.20, 55.86){\circle*{0.5}}
 \put( 36.30, 55.75){\circle*{0.5}}
 \put( 36.40, 55.63){\circle*{0.5}}
 \put( 36.50, 55.52){\circle*{0.5}}
 \put( 36.60, 55.40){\circle*{0.5}}
 \put( 36.70, 55.29){\circle*{0.5}}
 \put( 36.80, 55.18){\circle*{0.5}}
 \put( 36.90, 55.06){\circle*{0.5}}
 \put( 37.00, 54.95){\circle*{0.5}}
 \put( 37.10, 54.84){\circle*{0.5}}
 \put( 37.20, 54.73){\circle*{0.5}}
 \put( 37.30, 54.63){\circle*{0.5}}
 \put( 37.40, 54.52){\circle*{0.5}}
 \put( 37.50, 54.41){\circle*{0.5}}
 \put( 37.60, 54.31){\circle*{0.5}}
 \put( 37.70, 54.20){\circle*{0.5}}
 \put( 37.80, 54.10){\circle*{0.5}}
 \put( 37.90, 53.99){\circle*{0.5}}
 \put( 38.00, 53.89){\circle*{0.5}}
 \put( 38.10, 53.79){\circle*{0.5}}
 \put( 38.20, 53.68){\circle*{0.5}}
 \put( 38.30, 53.58){\circle*{0.5}}
 \put( 38.40, 53.48){\circle*{0.5}}
 \put( 38.50, 53.38){\circle*{0.5}}
 \put( 38.60, 53.28){\circle*{0.5}}
 \put( 38.70, 53.18){\circle*{0.5}}
 \put( 38.80, 53.09){\circle*{0.5}}
 \put( 38.90, 52.99){\circle*{0.5}}
 \put( 39.00, 52.89){\circle*{0.5}}
 \put( 39.10, 52.79){\circle*{0.5}}
 \put( 39.20, 52.70){\circle*{0.5}}
 \put( 39.30, 52.60){\circle*{0.5}}
 \put( 39.40, 52.51){\circle*{0.5}}
 \put( 39.50, 52.42){\circle*{0.5}}
 \put( 39.60, 52.32){\circle*{0.5}}
 \put( 39.70, 52.23){\circle*{0.5}}
 \put( 39.80, 52.14){\circle*{0.5}}
 \put( 39.90, 52.05){\circle*{0.5}}
 \put( 40.00, 51.96){\circle*{0.5}}
 \put( 40.10, 51.87){\circle*{0.5}}
 \put( 40.20, 51.78){\circle*{0.5}}
 \put( 40.30, 51.69){\circle*{0.5}}
 \put( 40.40, 51.60){\circle*{0.5}}
 \put( 40.50, 51.51){\circle*{0.5}}
 \put( 40.60, 51.42){\circle*{0.5}}
 \put( 40.70, 51.34){\circle*{0.5}}
 \put( 40.80, 51.25){\circle*{0.5}}
 \put( 40.90, 51.16){\circle*{0.5}}
 \put( 41.00, 51.08){\circle*{0.5}}
 \put( 41.10, 50.99){\circle*{0.5}}
 \put( 41.20, 50.91){\circle*{0.5}}
 \put( 41.30, 50.82){\circle*{0.5}}
 \put( 41.40, 50.74){\circle*{0.5}}
 \put( 41.50, 50.66){\circle*{0.5}}
 \put( 41.60, 50.58){\circle*{0.5}}
 \put( 41.70, 50.49){\circle*{0.5}}
 \put( 41.80, 50.41){\circle*{0.5}}
 \put( 41.90, 50.33){\circle*{0.5}}
 \put( 42.00, 50.25){\circle*{0.5}}
 \put( 42.10, 50.17){\circle*{0.5}}
 \put( 42.20, 50.09){\circle*{0.5}}
 \put( 42.30, 50.01){\circle*{0.5}}
 \put( 42.40, 49.93){\circle*{0.5}}
 \put( 42.50, 49.85){\circle*{0.5}}
 \put( 42.60, 49.78){\circle*{0.5}}
 \put( 42.70, 49.70){\circle*{0.5}}
 \put( 42.80, 49.62){\circle*{0.5}}
 \put( 42.90, 49.54){\circle*{0.5}}
 \put( 43.00, 49.47){\circle*{0.5}}
 \put( 43.10, 49.39){\circle*{0.5}}
 \put( 43.20, 49.32){\circle*{0.5}}
 \put( 43.30, 49.24){\circle*{0.5}}
 \put( 43.40, 49.17){\circle*{0.5}}
 \put( 43.50, 49.09){\circle*{0.5}}
 \put( 43.60, 49.02){\circle*{0.5}}
 \put( 43.70, 48.95){\circle*{0.5}}
 \put( 43.80, 48.87){\circle*{0.5}}
 \put( 43.90, 48.80){\circle*{0.5}}
 \put( 44.00, 48.73){\circle*{0.5}}
 \put( 44.10, 48.66){\circle*{0.5}}
 \put( 44.20, 48.58){\circle*{0.5}}
 \put( 44.30, 48.51){\circle*{0.5}}
 \put( 44.40, 48.44){\circle*{0.5}}
 \put( 44.50, 48.37){\circle*{0.5}}
 \put( 44.60, 48.30){\circle*{0.5}}
 \put( 44.70, 48.23){\circle*{0.5}}
 \put( 44.80, 48.16){\circle*{0.5}}
 \put( 44.90, 48.10){\circle*{0.5}}
 \put( 45.00, 48.03){\circle*{0.5}}
 \put( 45.10, 47.96){\circle*{0.5}}
 \put( 45.20, 47.89){\circle*{0.5}}
 \put( 45.30, 47.82){\circle*{0.5}}
 \put( 45.40, 47.76){\circle*{0.5}}
 \put( 45.50, 47.69){\circle*{0.5}}
 \put( 45.60, 47.62){\circle*{0.5}}
 \put( 45.70, 47.56){\circle*{0.5}}
 \put( 45.80, 47.49){\circle*{0.5}}
 \put( 45.90, 47.43){\circle*{0.5}}
 \put( 46.00, 47.36){\circle*{0.5}}
 \put( 46.10, 47.30){\circle*{0.5}}
 \put( 46.20, 47.23){\circle*{0.5}}
 \put( 46.30, 47.17){\circle*{0.5}}
 \put( 46.40, 47.10){\circle*{0.5}}
 \put( 46.50, 47.04){\circle*{0.5}}
 \put( 46.60, 46.98){\circle*{0.5}}
 \put( 46.70, 46.91){\circle*{0.5}}
 \put( 46.80, 46.85){\circle*{0.5}}
 \put( 46.90, 46.79){\circle*{0.5}}
 \put( 47.00, 46.73){\circle*{0.5}}
 \put( 47.10, 46.66){\circle*{0.5}}
 \put( 47.20, 46.60){\circle*{0.5}}
 \put( 47.30, 46.54){\circle*{0.5}}
 \put( 47.40, 46.48){\circle*{0.5}}
 \put( 47.50, 46.42){\circle*{0.5}}
 \put( 47.60, 46.36){\circle*{0.5}}
 \put( 47.70, 46.30){\circle*{0.5}}
 \put( 47.80, 46.24){\circle*{0.5}}
 \put( 47.90, 46.18){\circle*{0.5}}
 \put( 48.00, 46.12){\circle*{0.5}}
 \put( 48.10, 46.06){\circle*{0.5}}
 \put( 48.20, 46.01){\circle*{0.5}}
 \put( 48.30, 45.95){\circle*{0.5}}
 \put( 48.40, 45.89){\circle*{0.5}}
 \put( 48.50, 45.83){\circle*{0.5}}
 \put( 48.60, 45.77){\circle*{0.5}}
 \put( 48.70, 45.72){\circle*{0.5}}
 \put( 48.80, 45.66){\circle*{0.5}}
 \put( 48.90, 45.60){\circle*{0.5}}
 \put( 49.00, 45.55){\circle*{0.5}}
 \put( 49.10, 45.49){\circle*{0.5}}
 \put( 49.20, 45.43){\circle*{0.5}}
 \put( 49.30, 45.38){\circle*{0.5}}
 \put( 49.40, 45.32){\circle*{0.5}}
 \put( 49.50, 45.27){\circle*{0.5}}
 \put( 49.60, 45.21){\circle*{0.5}}
 \put( 49.70, 45.16){\circle*{0.5}}
 \put( 49.80, 45.10){\circle*{0.5}}
 \put( 49.90, 45.05){\circle*{0.5}}
 \put( 50.00, 45.00){\circle*{0.5}}
 \put( 50.10, 44.94){\circle*{0.5}}
 \put( 50.20, 44.89){\circle*{0.5}}
 \put( 50.30, 44.83){\circle*{0.5}}
 \put( 50.40, 44.78){\circle*{0.5}}
 \put( 50.50, 44.73){\circle*{0.5}}
 \put( 50.60, 44.68){\circle*{0.5}}
 \put( 50.70, 44.62){\circle*{0.5}}
 \put( 50.80, 44.57){\circle*{0.5}}
 \put( 50.90, 44.52){\circle*{0.5}}
 \put( 51.00, 44.47){\circle*{0.5}}
 \put( 51.10, 44.42){\circle*{0.5}}
 \put( 51.20, 44.37){\circle*{0.5}}
 \put( 51.30, 44.31){\circle*{0.5}}
 \put( 51.40, 44.26){\circle*{0.5}}
 \put( 51.50, 44.21){\circle*{0.5}}
 \put( 51.60, 44.16){\circle*{0.5}}
 \put( 51.70, 44.11){\circle*{0.5}}
 \put( 51.80, 44.06){\circle*{0.5}}
 \put( 51.90, 44.01){\circle*{0.5}}
 \put( 52.00, 43.96){\circle*{0.5}}
 \put( 52.10, 43.91){\circle*{0.5}}
 \put( 52.20, 43.87){\circle*{0.5}}
 \put( 52.30, 43.82){\circle*{0.5}}
 \put( 52.40, 43.77){\circle*{0.5}}
 \put( 52.50, 43.72){\circle*{0.5}}
 \put( 52.60, 43.67){\circle*{0.5}}
 \put( 52.70, 43.62){\circle*{0.5}}
 \put( 52.80, 43.57){\circle*{0.5}}
 \put( 52.90, 43.53){\circle*{0.5}}
 \put( 53.00, 43.48){\circle*{0.5}}
 \put( 53.10, 43.43){\circle*{0.5}}
 \put( 53.20, 43.38){\circle*{0.5}}
 \put( 53.30, 43.34){\circle*{0.5}}
 \put( 53.40, 43.29){\circle*{0.5}}
 \put( 53.50, 43.24){\circle*{0.5}}
 \put( 53.60, 43.20){\circle*{0.5}}
 \put( 53.70, 43.15){\circle*{0.5}}
 \put( 53.80, 43.11){\circle*{0.5}}
 \put( 53.90, 43.06){\circle*{0.5}}
 \put( 54.00, 43.01){\circle*{0.5}}
 \put( 54.10, 42.97){\circle*{0.5}}
 \put( 54.20, 42.92){\circle*{0.5}}
 \put( 54.30, 42.88){\circle*{0.5}}
 \put( 54.40, 42.83){\circle*{0.5}}
 \put( 54.50, 42.79){\circle*{0.5}}
 \put( 54.60, 42.74){\circle*{0.5}}
 \put( 54.70, 42.70){\circle*{0.5}}
 \put( 54.80, 42.66){\circle*{0.5}}
 \put( 54.90, 42.61){\circle*{0.5}}
 \put( 55.00, 42.57){\circle*{0.5}}
 \put( 55.10, 42.52){\circle*{0.5}}
 \put( 55.20, 42.48){\circle*{0.5}}
 \put( 55.30, 42.44){\circle*{0.5}}
 \put( 55.40, 42.39){\circle*{0.5}}
 \put( 55.50, 42.35){\circle*{0.5}}
 \put( 55.60, 42.31){\circle*{0.5}}
 \put( 55.70, 42.26){\circle*{0.5}}
 \put( 55.80, 42.22){\circle*{0.5}}
 \put( 55.90, 42.18){\circle*{0.5}}
 \put( 56.00, 42.14){\circle*{0.5}}
 \put( 56.10, 42.10){\circle*{0.5}}
 \put( 56.20, 42.05){\circle*{0.5}}
 \put( 56.30, 42.01){\circle*{0.5}}
 \put( 56.40, 41.97){\circle*{0.5}}
 \put( 56.50, 41.93){\circle*{0.5}}
 \put( 56.60, 41.89){\circle*{0.5}}
 \put( 56.70, 41.85){\circle*{0.5}}
 \put( 56.80, 41.81){\circle*{0.5}}
 \put( 56.90, 41.76){\circle*{0.5}}
 \put( 57.00, 41.72){\circle*{0.5}}
 \put( 57.10, 41.68){\circle*{0.5}}
 \put( 57.20, 41.64){\circle*{0.5}}
 \put( 57.30, 41.60){\circle*{0.5}}
 \put( 57.40, 41.56){\circle*{0.5}}
 \put( 57.50, 41.52){\circle*{0.5}}
 \put( 57.60, 41.48){\circle*{0.5}}
 \put( 57.70, 41.44){\circle*{0.5}}
 \put( 57.80, 41.40){\circle*{0.5}}
 \put( 57.90, 41.36){\circle*{0.5}}
 \put( 58.00, 41.32){\circle*{0.5}}
 \put( 58.10, 41.29){\circle*{0.5}}
 \put( 58.20, 41.25){\circle*{0.5}}
 \put( 58.30, 41.21){\circle*{0.5}}
 \put( 58.40, 41.17){\circle*{0.5}}
 \put( 58.50, 41.13){\circle*{0.5}}
 \put( 58.60, 41.09){\circle*{0.5}}
 \put( 58.70, 41.05){\circle*{0.5}}
 \put( 58.80, 41.02){\circle*{0.5}}
 \put( 58.90, 40.98){\circle*{0.5}}
 \put( 59.00, 40.94){\circle*{0.5}}
 \put( 59.10, 40.90){\circle*{0.5}}
 \put( 59.20, 40.86){\circle*{0.5}}
 \put( 59.30, 40.83){\circle*{0.5}}
 \put( 59.40, 40.79){\circle*{0.5}}
 \put( 59.50, 40.75){\circle*{0.5}}
 \put( 59.60, 40.71){\circle*{0.5}}
 \put( 59.70, 40.68){\circle*{0.5}}
 \put( 59.80, 40.64){\circle*{0.5}}
 \put( 59.90, 40.60){\circle*{0.5}}
 \put( 60.00, 40.57){\circle*{0.5}}
 \put( 60.10, 40.53){\circle*{0.5}}
 \put( 60.20, 40.49){\circle*{0.5}}
 \put( 60.30, 40.46){\circle*{0.5}}
 \put( 60.40, 40.42){\circle*{0.5}}
 \put( 60.50, 40.39){\circle*{0.5}}
 \put( 60.60, 40.35){\circle*{0.5}}
 \put( 60.70, 40.31){\circle*{0.5}}
 \put( 60.80, 40.28){\circle*{0.5}}
 \put( 60.90, 40.24){\circle*{0.5}}
 \put( 61.00, 40.21){\circle*{0.5}}
 \put( 61.10, 40.17){\circle*{0.5}}
 \put( 61.20, 40.14){\circle*{0.5}}
 \put( 61.30, 40.10){\circle*{0.5}}
 \put( 61.40, 40.07){\circle*{0.5}}
 \put( 61.50, 40.03){\circle*{0.5}}
 \put( 61.60, 40.00){\circle*{0.5}}
 \put( 61.70, 39.96){\circle*{0.5}}
 \put( 61.80, 39.93){\circle*{0.5}}
 \put( 61.90, 39.90){\circle*{0.5}}
 \put( 62.00, 39.86){\circle*{0.5}}
 \put( 62.10, 39.83){\circle*{0.5}}
 \put( 62.20, 39.79){\circle*{0.5}}
 \put( 62.30, 39.76){\circle*{0.5}}
 \put( 62.40, 39.73){\circle*{0.5}}
 \put( 62.50, 39.69){\circle*{0.5}}
 \put( 62.60, 39.66){\circle*{0.5}}
 \put( 62.70, 39.62){\circle*{0.5}}
 \put( 62.80, 39.59){\circle*{0.5}}
 \put( 62.90, 39.56){\circle*{0.5}}
 \put( 63.00, 39.52){\circle*{0.5}}
 \put( 63.10, 39.49){\circle*{0.5}}
 \put( 63.20, 39.46){\circle*{0.5}}
 \put( 63.30, 39.43){\circle*{0.5}}
 \put( 63.40, 39.39){\circle*{0.5}}
 \put( 63.50, 39.36){\circle*{0.5}}
 \put( 63.60, 39.33){\circle*{0.5}}
 \put( 63.70, 39.30){\circle*{0.5}}
 \put( 63.80, 39.26){\circle*{0.5}}
 \put( 63.90, 39.23){\circle*{0.5}}
 \put( 64.00, 39.20){\circle*{0.5}}
 \put( 64.10, 39.17){\circle*{0.5}}
 \put( 64.20, 39.14){\circle*{0.5}}
 \put( 64.30, 39.10){\circle*{0.5}}
 \put( 64.40, 39.07){\circle*{0.5}}
 \put( 64.50, 39.04){\circle*{0.5}}
 \put( 64.60, 39.01){\circle*{0.5}}
 \put( 64.70, 38.98){\circle*{0.5}}
 \put( 64.80, 38.95){\circle*{0.5}}
 \put( 64.90, 38.91){\circle*{0.5}}
 \put( 65.00, 38.88){\circle*{0.5}}
 \put( 65.10, 38.85){\circle*{0.5}}
 \put( 65.20, 38.82){\circle*{0.5}}
 \put( 65.30, 38.79){\circle*{0.5}}
 \put( 65.40, 38.76){\circle*{0.5}}
 \put( 65.50, 38.73){\circle*{0.5}}
 \put( 65.60, 38.70){\circle*{0.5}}
 \put( 65.70, 38.67){\circle*{0.5}}
 \put( 65.80, 38.64){\circle*{0.5}}
 \put( 65.90, 38.61){\circle*{0.5}}
 \put( 66.00, 38.58){\circle*{0.5}}
 \put( 66.10, 38.55){\circle*{0.5}}
 \put( 66.20, 38.52){\circle*{0.5}}
 \put( 66.30, 38.49){\circle*{0.5}}
 \put( 66.40, 38.46){\circle*{0.5}}
 \put( 66.50, 38.43){\circle*{0.5}}
 \put( 66.60, 38.40){\circle*{0.5}}
 \put( 66.70, 38.37){\circle*{0.5}}
 \put( 66.80, 38.34){\circle*{0.5}}
 \put( 66.90, 38.31){\circle*{0.5}}
 \put( 67.00, 38.28){\circle*{0.5}}
 \put( 67.10, 38.25){\circle*{0.5}}
 \put( 67.20, 38.22){\circle*{0.5}}
 \put( 67.30, 38.19){\circle*{0.5}}
 \put( 67.40, 38.16){\circle*{0.5}}
 \put( 67.50, 38.14){\circle*{0.5}}
 \put( 67.60, 38.11){\circle*{0.5}}
 \put( 67.70, 38.08){\circle*{0.5}}
 \put( 67.80, 38.05){\circle*{0.5}}
 \put( 67.90, 38.02){\circle*{0.5}}
 \put( 68.00, 37.99){\circle*{0.5}}
 \put( 68.10, 37.96){\circle*{0.5}}
 \put( 68.20, 37.94){\circle*{0.5}}
 \put( 68.30, 37.91){\circle*{0.5}}
 \put( 68.40, 37.88){\circle*{0.5}}
 \put( 68.50, 37.85){\circle*{0.5}}
 \put( 68.60, 37.82){\circle*{0.5}}
 \put( 68.70, 37.80){\circle*{0.5}}
 \put( 68.80, 37.77){\circle*{0.5}}
 \put( 68.90, 37.74){\circle*{0.5}}
 \put( 69.00, 37.71){\circle*{0.5}}
 \put( 69.10, 37.69){\circle*{0.5}}
 \put( 69.20, 37.66){\circle*{0.5}}
 \put( 69.30, 37.63){\circle*{0.5}}
 \put( 69.40, 37.60){\circle*{0.5}}
 \put( 69.50, 37.58){\circle*{0.5}}
 \put( 69.60, 37.55){\circle*{0.5}}
 \put( 69.70, 37.52){\circle*{0.5}}
 \put( 69.80, 37.50){\circle*{0.5}}
 \put( 69.90, 37.47){\circle*{0.5}}
 \put( 70.00, 37.44){\circle*{0.5}}
 \put( 70.10, 37.41){\circle*{0.5}}
 \put( 70.20, 37.39){\circle*{0.5}}
 \put( 70.30, 37.36){\circle*{0.5}}
 \put( 70.40, 37.33){\circle*{0.5}}
 \put( 70.50, 37.31){\circle*{0.5}}
 \put( 70.60, 37.28){\circle*{0.5}}
 \put( 70.70, 37.26){\circle*{0.5}}
 \put( 70.80, 37.23){\circle*{0.5}}
 \put( 70.90, 37.20){\circle*{0.5}}
 \put( 71.00, 37.18){\circle*{0.5}}
 \put( 71.10, 37.15){\circle*{0.5}}
 \put( 71.20, 37.12){\circle*{0.5}}
 \put( 71.30, 37.10){\circle*{0.5}}
 \put( 71.40, 37.07){\circle*{0.5}}
 \put( 71.50, 37.05){\circle*{0.5}}
 \put( 71.60, 37.02){\circle*{0.5}}
 \put( 71.70, 37.00){\circle*{0.5}}
 \put( 71.80, 36.97){\circle*{0.5}}
 \put( 71.90, 36.95){\circle*{0.5}}
 \put( 72.00, 36.92){\circle*{0.5}}
 \put( 72.10, 36.89){\circle*{0.5}}
 \put( 72.20, 36.87){\circle*{0.5}}
 \put( 72.30, 36.84){\circle*{0.5}}
 \put( 72.40, 36.82){\circle*{0.5}}
 \put( 72.50, 36.79){\circle*{0.5}}
 \put( 72.60, 36.77){\circle*{0.5}}
 \put( 72.70, 36.74){\circle*{0.5}}
 \put( 72.80, 36.72){\circle*{0.5}}
 \put( 72.90, 36.69){\circle*{0.5}}
 \put( 73.00, 36.67){\circle*{0.5}}
 \put( 73.10, 36.64){\circle*{0.5}}
 \put( 73.20, 36.62){\circle*{0.5}}
 \put( 73.30, 36.60){\circle*{0.5}}
 \put( 73.40, 36.57){\circle*{0.5}}
 \put( 73.50, 36.55){\circle*{0.5}}
 \put( 73.60, 36.52){\circle*{0.5}}
 \put( 73.70, 36.50){\circle*{0.5}}
 \put( 73.80, 36.47){\circle*{0.5}}
 \put( 73.90, 36.45){\circle*{0.5}}
 \put( 74.00, 36.43){\circle*{0.5}}
 \put( 74.10, 36.40){\circle*{0.5}}
 \put( 74.20, 36.38){\circle*{0.5}}
 \put( 74.30, 36.35){\circle*{0.5}}
 \put( 74.40, 36.33){\circle*{0.5}}
 \put( 74.50, 36.31){\circle*{0.5}}
 \put( 74.60, 36.28){\circle*{0.5}}
 \put( 74.70, 36.26){\circle*{0.5}}
 \put( 74.80, 36.24){\circle*{0.5}}
 \put( 74.90, 36.21){\circle*{0.5}}
 \put( 75.00, 36.19){\circle*{0.5}}
 \put( 75.10, 36.16){\circle*{0.5}}
 \put( 75.20, 36.14){\circle*{0.5}}
 \put( 75.30, 36.12){\circle*{0.5}}
 \put( 75.40, 36.09){\circle*{0.5}}
 \put( 75.50, 36.07){\circle*{0.5}}
 \put( 75.60, 36.05){\circle*{0.5}}
 \put( 75.70, 36.03){\circle*{0.5}}
 \put( 75.80, 36.00){\circle*{0.5}}
 \put( 75.90, 35.98){\circle*{0.5}}
 \put( 76.00, 35.96){\circle*{0.5}}
 \put( 76.10, 35.93){\circle*{0.5}}
 \put( 76.20, 35.91){\circle*{0.5}}
 \put( 76.30, 35.89){\circle*{0.5}}
 \put( 76.40, 35.87){\circle*{0.5}}
 \put( 76.50, 35.84){\circle*{0.5}}
 \put( 76.60, 35.82){\circle*{0.5}}
 \put( 76.70, 35.80){\circle*{0.5}}
 \put( 76.80, 35.78){\circle*{0.5}}
 \put( 76.90, 35.75){\circle*{0.5}}
 \put( 77.00, 35.73){\circle*{0.5}}
 \put( 77.10, 35.71){\circle*{0.5}}
 \put( 77.20, 35.69){\circle*{0.5}}
 \put( 77.30, 35.66){\circle*{0.5}}
 \put( 77.40, 35.64){\circle*{0.5}}
 \put( 77.50, 35.62){\circle*{0.5}}
 \put( 77.60, 35.60){\circle*{0.5}}
 \put( 77.70, 35.58){\circle*{0.5}}
 \put( 77.80, 35.55){\circle*{0.5}}
 \put( 77.90, 35.53){\circle*{0.5}}
 \put( 78.00, 35.51){\circle*{0.5}}
 \put( 78.10, 35.49){\circle*{0.5}}
 \put( 78.20, 35.47){\circle*{0.5}}
 \put( 78.30, 35.45){\circle*{0.5}}
 \put( 78.40, 35.42){\circle*{0.5}}
 \put( 78.50, 35.40){\circle*{0.5}}
 \put( 78.60, 35.38){\circle*{0.5}}
 \put( 78.70, 35.36){\circle*{0.5}}
 \put( 78.80, 35.34){\circle*{0.5}}
 \put( 78.90, 35.32){\circle*{0.5}}
 \put( 79.00, 35.30){\circle*{0.5}}
 \put( 79.10, 35.27){\circle*{0.5}}
 \put( 79.20, 35.25){\circle*{0.5}}
 \put( 79.30, 35.23){\circle*{0.5}}
 \put( 79.40, 35.21){\circle*{0.5}}
 \put( 79.50, 35.19){\circle*{0.5}}
 \put( 79.60, 35.17){\circle*{0.5}}
 \put( 79.70, 35.15){\circle*{0.5}}
 \put( 79.80, 35.13){\circle*{0.5}}
 \put( 79.90, 35.11){\circle*{0.5}}
 \put( 80.00, 35.09){\circle*{0.5}}

\put(81,36){$\Upsilon(nS)$}

\put(25,51.0){\circle*{1.6}}
\put(35,38.3){\circle*{1.6}}
\put(45,30.5){\circle*{1.6}}
\put(55,28.0){\circle*{1.6}}
\put(65,24.5){\circle*{1.6}}
\put(55,25.0){\line(0,1){6}}


 \put( 25.10, 50.74){\circle*{0.5}}
 \put( 25.20, 50.49){\circle*{0.5}}
 \put( 25.30, 50.24){\circle*{0.5}}
 \put( 25.40, 50.00){\circle*{0.5}}
 \put( 25.50, 49.76){\circle*{0.5}}
 \put( 25.60, 49.52){\circle*{0.5}}
 \put( 25.70, 49.29){\circle*{0.5}}
 \put( 25.80, 49.06){\circle*{0.5}}
 \put( 25.90, 48.84){\circle*{0.5}}
 \put( 26.00, 48.62){\circle*{0.5}}
 \put( 26.10, 48.40){\circle*{0.5}}
 \put( 26.20, 48.18){\circle*{0.5}}
 \put( 26.30, 47.97){\circle*{0.5}}
 \put( 26.40, 47.76){\circle*{0.5}}
 \put( 26.50, 47.55){\circle*{0.5}}
 \put( 26.60, 47.35){\circle*{0.5}}
 \put( 26.70, 47.15){\circle*{0.5}}
 \put( 26.80, 46.96){\circle*{0.5}}
 \put( 26.90, 46.76){\circle*{0.5}}
 \put( 27.00, 46.57){\circle*{0.5}}
 \put( 27.10, 46.38){\circle*{0.5}}
 \put( 27.20, 46.19){\circle*{0.5}}
 \put( 27.30, 46.01){\circle*{0.5}}
 \put( 27.40, 45.83){\circle*{0.5}}
 \put( 27.50, 45.65){\circle*{0.5}}
 \put( 27.60, 45.47){\circle*{0.5}}
 \put( 27.70, 45.30){\circle*{0.5}}
 \put( 27.80, 45.13){\circle*{0.5}}
 \put( 27.90, 44.96){\circle*{0.5}}
 \put( 28.00, 44.79){\circle*{0.5}}
 \put( 28.10, 44.62){\circle*{0.5}}
 \put( 28.20, 44.46){\circle*{0.5}}
 \put( 28.30, 44.30){\circle*{0.5}}
 \put( 28.40, 44.14){\circle*{0.5}}
 \put( 28.50, 43.98){\circle*{0.5}}
 \put( 28.60, 43.83){\circle*{0.5}}
 \put( 28.70, 43.67){\circle*{0.5}}
 \put( 28.80, 43.52){\circle*{0.5}}
 \put( 28.90, 43.37){\circle*{0.5}}
 \put( 29.00, 43.23){\circle*{0.5}}
 \put( 29.10, 43.08){\circle*{0.5}}
 \put( 29.20, 42.93){\circle*{0.5}}
 \put( 29.30, 42.79){\circle*{0.5}}
 \put( 29.40, 42.65){\circle*{0.5}}
 \put( 29.50, 42.51){\circle*{0.5}}
 \put( 29.60, 42.37){\circle*{0.5}}
 \put( 29.70, 42.24){\circle*{0.5}}
 \put( 29.80, 42.10){\circle*{0.5}}
 \put( 29.90, 41.97){\circle*{0.5}}
 \put( 30.00, 41.84){\circle*{0.5}}
 \put( 30.10, 41.71){\circle*{0.5}}
 \put( 30.20, 41.58){\circle*{0.5}}
 \put( 30.30, 41.45){\circle*{0.5}}
 \put( 30.40, 41.32){\circle*{0.5}}
 \put( 30.50, 41.20){\circle*{0.5}}
 \put( 30.60, 41.08){\circle*{0.5}}
 \put( 30.70, 40.95){\circle*{0.5}}
 \put( 30.80, 40.83){\circle*{0.5}}
 \put( 30.90, 40.71){\circle*{0.5}}
 \put( 31.00, 40.59){\circle*{0.5}}
 \put( 31.10, 40.48){\circle*{0.5}}
 \put( 31.20, 40.36){\circle*{0.5}}
 \put( 31.30, 40.25){\circle*{0.5}}
 \put( 31.40, 40.13){\circle*{0.5}}
 \put( 31.50, 40.02){\circle*{0.5}}
 \put( 31.60, 39.91){\circle*{0.5}}
 \put( 31.70, 39.80){\circle*{0.5}}
 \put( 31.80, 39.69){\circle*{0.5}}
 \put( 31.90, 39.58){\circle*{0.5}}
 \put( 32.00, 39.47){\circle*{0.5}}
 \put( 32.10, 39.37){\circle*{0.5}}
 \put( 32.20, 39.26){\circle*{0.5}}
 \put( 32.30, 39.16){\circle*{0.5}}
 \put( 32.40, 39.06){\circle*{0.5}}
 \put( 32.50, 38.95){\circle*{0.5}}
 \put( 32.60, 38.85){\circle*{0.5}}
 \put( 32.70, 38.75){\circle*{0.5}}
 \put( 32.80, 38.65){\circle*{0.5}}
 \put( 32.90, 38.56){\circle*{0.5}}
 \put( 33.00, 38.46){\circle*{0.5}}
 \put( 33.10, 38.36){\circle*{0.5}}
 \put( 33.20, 38.27){\circle*{0.5}}
 \put( 33.30, 38.17){\circle*{0.5}}
 \put( 33.40, 38.08){\circle*{0.5}}
 \put( 33.50, 37.98){\circle*{0.5}}
 \put( 33.60, 37.89){\circle*{0.5}}
 \put( 33.70, 37.80){\circle*{0.5}}
 \put( 33.80, 37.71){\circle*{0.5}}
 \put( 33.90, 37.62){\circle*{0.5}}
 \put( 34.00, 37.53){\circle*{0.5}}
 \put( 34.10, 37.44){\circle*{0.5}}
 \put( 34.20, 37.36){\circle*{0.5}}
 \put( 34.30, 37.27){\circle*{0.5}}
 \put( 34.40, 37.18){\circle*{0.5}}
 \put( 34.50, 37.10){\circle*{0.5}}
 \put( 34.60, 37.01){\circle*{0.5}}
 \put( 34.70, 36.93){\circle*{0.5}}
 \put( 34.80, 36.85){\circle*{0.5}}
 \put( 34.90, 36.76){\circle*{0.5}}
 \put( 35.00, 36.68){\circle*{0.5}}
 \put( 35.10, 36.60){\circle*{0.5}}
 \put( 35.20, 36.52){\circle*{0.5}}
 \put( 35.30, 36.44){\circle*{0.5}}
 \put( 35.40, 36.36){\circle*{0.5}}
 \put( 35.50, 36.28){\circle*{0.5}}
 \put( 35.60, 36.20){\circle*{0.5}}
 \put( 35.70, 36.13){\circle*{0.5}}
 \put( 35.80, 36.05){\circle*{0.5}}
 \put( 35.90, 35.97){\circle*{0.5}}
 \put( 36.00, 35.90){\circle*{0.5}}
 \put( 36.10, 35.82){\circle*{0.5}}
 \put( 36.20, 35.75){\circle*{0.5}}
 \put( 36.30, 35.68){\circle*{0.5}}
 \put( 36.40, 35.60){\circle*{0.5}}
 \put( 36.50, 35.53){\circle*{0.5}}
 \put( 36.60, 35.46){\circle*{0.5}}
 \put( 36.70, 35.39){\circle*{0.5}}
 \put( 36.80, 35.32){\circle*{0.5}}
 \put( 36.90, 35.25){\circle*{0.5}}
 \put( 37.00, 35.18){\circle*{0.5}}
 \put( 37.10, 35.11){\circle*{0.5}}
 \put( 37.20, 35.04){\circle*{0.5}}
 \put( 37.30, 34.97){\circle*{0.5}}
 \put( 37.40, 34.90){\circle*{0.5}}
 \put( 37.50, 34.83){\circle*{0.5}}
 \put( 37.60, 34.77){\circle*{0.5}}
 \put( 37.70, 34.70){\circle*{0.5}}
 \put( 37.80, 34.64){\circle*{0.5}}
 \put( 37.90, 34.57){\circle*{0.5}}
 \put( 38.00, 34.51){\circle*{0.5}}
 \put( 38.10, 34.44){\circle*{0.5}}
 \put( 38.20, 34.38){\circle*{0.5}}
 \put( 38.30, 34.31){\circle*{0.5}}
 \put( 38.40, 34.25){\circle*{0.5}}
 \put( 38.50, 34.19){\circle*{0.5}}
 \put( 38.60, 34.13){\circle*{0.5}}
 \put( 38.70, 34.06){\circle*{0.5}}
 \put( 38.80, 34.00){\circle*{0.5}}
 \put( 38.90, 33.94){\circle*{0.5}}
 \put( 39.00, 33.88){\circle*{0.5}}
 \put( 39.10, 33.82){\circle*{0.5}}
 \put( 39.20, 33.76){\circle*{0.5}}
 \put( 39.30, 33.70){\circle*{0.5}}
 \put( 39.40, 33.64){\circle*{0.5}}
 \put( 39.50, 33.58){\circle*{0.5}}
 \put( 39.60, 33.53){\circle*{0.5}}
 \put( 39.70, 33.47){\circle*{0.5}}
 \put( 39.80, 33.41){\circle*{0.5}}
 \put( 39.90, 33.35){\circle*{0.5}}
 \put( 40.00, 33.30){\circle*{0.5}}
 \put( 40.10, 33.24){\circle*{0.5}}
 \put( 40.20, 33.19){\circle*{0.5}}
 \put( 40.30, 33.13){\circle*{0.5}}
 \put( 40.40, 33.08){\circle*{0.5}}
 \put( 40.50, 33.02){\circle*{0.5}}
 \put( 40.60, 32.97){\circle*{0.5}}
 \put( 40.70, 32.91){\circle*{0.5}}
 \put( 40.80, 32.86){\circle*{0.5}}
 \put( 40.90, 32.80){\circle*{0.5}}
 \put( 41.00, 32.75){\circle*{0.5}}
 \put( 41.10, 32.70){\circle*{0.5}}
 \put( 41.20, 32.65){\circle*{0.5}}
 \put( 41.30, 32.59){\circle*{0.5}}
 \put( 41.40, 32.54){\circle*{0.5}}
 \put( 41.50, 32.49){\circle*{0.5}}
 \put( 41.60, 32.44){\circle*{0.5}}
 \put( 41.70, 32.39){\circle*{0.5}}
 \put( 41.80, 32.34){\circle*{0.5}}
 \put( 41.90, 32.29){\circle*{0.5}}
 \put( 42.00, 32.24){\circle*{0.5}}
 \put( 42.10, 32.19){\circle*{0.5}}
 \put( 42.20, 32.14){\circle*{0.5}}
 \put( 42.30, 32.09){\circle*{0.5}}
 \put( 42.40, 32.04){\circle*{0.5}}
 \put( 42.50, 31.99){\circle*{0.5}}
 \put( 42.60, 31.95){\circle*{0.5}}
 \put( 42.70, 31.90){\circle*{0.5}}
 \put( 42.80, 31.85){\circle*{0.5}}
 \put( 42.90, 31.80){\circle*{0.5}}
 \put( 43.00, 31.76){\circle*{0.5}}
 \put( 43.10, 31.71){\circle*{0.5}}
 \put( 43.20, 31.66){\circle*{0.5}}
 \put( 43.30, 31.62){\circle*{0.5}}
 \put( 43.40, 31.57){\circle*{0.5}}
 \put( 43.50, 31.53){\circle*{0.5}}
 \put( 43.60, 31.48){\circle*{0.5}}
 \put( 43.70, 31.44){\circle*{0.5}}
 \put( 43.80, 31.39){\circle*{0.5}}
 \put( 43.90, 31.35){\circle*{0.5}}
 \put( 44.00, 31.30){\circle*{0.5}}
 \put( 44.10, 31.26){\circle*{0.5}}
 \put( 44.20, 31.21){\circle*{0.5}}
 \put( 44.30, 31.17){\circle*{0.5}}
 \put( 44.40, 31.13){\circle*{0.5}}
 \put( 44.50, 31.08){\circle*{0.5}}
 \put( 44.60, 31.04){\circle*{0.5}}
 \put( 44.70, 31.00){\circle*{0.5}}
 \put( 44.80, 30.96){\circle*{0.5}}
 \put( 44.90, 30.91){\circle*{0.5}}
 \put( 45.00, 30.87){\circle*{0.5}}
 \put( 45.10, 30.83){\circle*{0.5}}
 \put( 45.20, 30.79){\circle*{0.5}}
 \put( 45.30, 30.5){\circle*{0.5}}
 \put( 45.40, 30.71){\circle*{0.5}}
 \put( 45.50, 30.67){\circle*{0.5}}
 \put( 45.60, 30.62){\circle*{0.5}}
 \put( 45.70, 30.58){\circle*{0.5}}
 \put( 45.80, 30.54){\circle*{0.5}}
 \put( 45.90, 30.50){\circle*{0.5}}
 \put( 46.00, 30.46){\circle*{0.5}}
 \put( 46.10, 30.42){\circle*{0.5}}
 \put( 46.20, 30.39){\circle*{0.5}}
 \put( 46.30, 30.35){\circle*{0.5}}
 \put( 46.40, 30.31){\circle*{0.5}}
 \put( 46.50, 30.27){\circle*{0.5}}
 \put( 46.60, 30.23){\circle*{0.5}}
 \put( 46.70, 30.19){\circle*{0.5}}
 \put( 46.80, 30.15){\circle*{0.5}}
 \put( 46.90, 30.12){\circle*{0.5}}
 \put( 47.00, 30.08){\circle*{0.5}}
 \put( 47.10, 30.04){\circle*{0.5}}
 \put( 47.20, 30.00){\circle*{0.5}}
 \put( 47.30, 29.97){\circle*{0.5}}
 \put( 47.40, 29.93){\circle*{0.5}}
 \put( 47.50, 29.89){\circle*{0.5}}
 \put( 47.60, 29.86){\circle*{0.5}}
 \put( 47.70, 29.82){\circle*{0.5}}
 \put( 47.80, 29.78){\circle*{0.5}}
 \put( 47.90, 29.75){\circle*{0.5}}
 \put( 48.00, 29.71){\circle*{0.5}}
 \put( 48.10, 29.68){\circle*{0.5}}
 \put( 48.20, 29.64){\circle*{0.5}}
 \put( 48.30, 29.60){\circle*{0.5}}
 \put( 48.40, 29.57){\circle*{0.5}}
 \put( 48.50, 29.53){\circle*{0.5}}
 \put( 48.60, 29.50){\circle*{0.5}}
 \put( 48.70, 29.46){\circle*{0.5}}
 \put( 48.80, 29.43){\circle*{0.5}}
 \put( 48.90, 29.40){\circle*{0.5}}
 \put( 49.00, 29.36){\circle*{0.5}}
 \put( 49.10, 29.33){\circle*{0.5}}
 \put( 49.20, 29.29){\circle*{0.5}}
 \put( 49.30, 29.26){\circle*{0.5}}
 \put( 49.40, 29.23){\circle*{0.5}}
 \put( 49.50, 29.19){\circle*{0.5}}
 \put( 49.60, 29.16){\circle*{0.5}}
 \put( 49.70, 29.13){\circle*{0.5}}
 \put( 49.80, 29.09){\circle*{0.5}}
 \put( 49.90, 29.06){\circle*{0.5}}
 \put( 50.00, 29.03){\circle*{0.5}}
 \put( 50.10, 29.00){\circle*{0.5}}
 \put( 50.20, 28.96){\circle*{0.5}}
 \put( 50.30, 28.93){\circle*{0.5}}
 \put( 50.40, 28.90){\circle*{0.5}}
 \put( 50.50, 28.87){\circle*{0.5}}
 \put( 50.60, 28.84){\circle*{0.5}}
 \put( 50.70, 28.80){\circle*{0.5}}
 \put( 50.80, 28.77){\circle*{0.5}}
 \put( 50.90, 28.74){\circle*{0.5}}
 \put( 51.00, 28.71){\circle*{0.5}}
 \put( 51.10, 28.68){\circle*{0.5}}
 \put( 51.20, 28.65){\circle*{0.5}}
 \put( 51.30, 28.62){\circle*{0.5}}
 \put( 51.40, 28.59){\circle*{0.5}}
 \put( 51.50, 28.56){\circle*{0.5}}
 \put( 51.60, 28.53){\circle*{0.5}}
 \put( 51.70, 28.50){\circle*{0.5}}
 \put( 51.80, 28.47){\circle*{0.5}}
 \put( 51.90, 28.44){\circle*{0.5}}
 \put( 52.00, 28.41){\circle*{0.5}}
 \put( 52.10, 28.38){\circle*{0.5}}
 \put( 52.20, 28.35){\circle*{0.5}}
 \put( 52.30, 28.32){\circle*{0.5}}
 \put( 52.40, 28.29){\circle*{0.5}}
 \put( 52.50, 28.26){\circle*{0.5}}
 \put( 52.60, 28.23){\circle*{0.5}}
 \put( 52.70, 28.20){\circle*{0.5}}
 \put( 52.80, 28.17){\circle*{0.5}}
 \put( 52.90, 28.14){\circle*{0.5}}
 \put( 53.00, 28.12){\circle*{0.5}}
 \put( 53.10, 28.09){\circle*{0.5}}
 \put( 53.20, 28.06){\circle*{0.5}}
 \put( 53.30, 28.03){\circle*{0.5}}
 \put( 53.40, 28.00){\circle*{0.5}}
 \put( 53.50, 27.98){\circle*{0.5}}
 \put( 53.60, 27.95){\circle*{0.5}}
 \put( 53.70, 27.92){\circle*{0.5}}
 \put( 53.80, 27.89){\circle*{0.5}}
 \put( 53.90, 27.86){\circle*{0.5}}
 \put( 54.00, 27.84){\circle*{0.5}}
 \put( 54.10, 27.81){\circle*{0.5}}
 \put( 54.20, 27.78){\circle*{0.5}}
 \put( 54.30, 27.76){\circle*{0.5}}
 \put( 54.40, 27.73){\circle*{0.5}}
 \put( 54.50, 27.70){\circle*{0.5}}
 \put( 54.60, 27.68){\circle*{0.5}}
 \put( 54.70, 27.65){\circle*{0.5}}
 \put( 54.80, 27.62){\circle*{0.5}}
 \put( 54.90, 27.60){\circle*{0.5}}
 \put( 55.00, 27.57){\circle*{0.5}}
 \put( 55.10, 27.54){\circle*{0.5}}
 \put( 55.20, 27.52){\circle*{0.5}}
 \put( 55.30, 27.49){\circle*{0.5}}
 \put( 55.40, 27.47){\circle*{0.5}}
 \put( 55.50, 27.44){\circle*{0.5}}
 \put( 55.60, 27.42){\circle*{0.5}}
 \put( 55.70, 27.39){\circle*{0.5}}
 \put( 55.80, 27.36){\circle*{0.5}}
 \put( 55.90, 27.34){\circle*{0.5}}
 \put( 56.00, 27.31){\circle*{0.5}}
 \put( 56.10, 27.29){\circle*{0.5}}
 \put( 56.20, 27.26){\circle*{0.5}}
 \put( 56.30, 27.24){\circle*{0.5}}
 \put( 56.40, 27.21){\circle*{0.5}}
 \put( 56.50, 27.19){\circle*{0.5}}
 \put( 56.60, 27.17){\circle*{0.5}}
 \put( 56.70, 27.14){\circle*{0.5}}
 \put( 56.80, 27.12){\circle*{0.5}}
 \put( 56.90, 27.09){\circle*{0.5}}
 \put( 57.00, 27.07){\circle*{0.5}}
 \put( 57.10, 27.04){\circle*{0.5}}
 \put( 57.20, 27.02){\circle*{0.5}}
 \put( 57.30, 27.00){\circle*{0.5}}
 \put( 57.40, 26.97){\circle*{0.5}}
 \put( 57.50, 26.95){\circle*{0.5}}
 \put( 57.60, 26.92){\circle*{0.5}}
 \put( 57.70, 26.90){\circle*{0.5}}
 \put( 57.80, 26.88){\circle*{0.5}}
 \put( 57.90, 26.85){\circle*{0.5}}
 \put( 58.00, 26.83){\circle*{0.5}}
 \put( 58.10, 26.81){\circle*{0.5}}
 \put( 58.20, 26.78){\circle*{0.5}}
 \put( 58.30, 26.76){\circle*{0.5}}
 \put( 58.40, 26.74){\circle*{0.5}}
 \put( 58.50, 26.72){\circle*{0.5}}
 \put( 58.60, 26.69){\circle*{0.5}}
 \put( 58.70, 26.67){\circle*{0.5}}
 \put( 58.80, 26.65){\circle*{0.5}}
 \put( 58.90, 26.62){\circle*{0.5}}
 \put( 59.00, 26.60){\circle*{0.5}}
 \put( 59.10, 26.58){\circle*{0.5}}
 \put( 59.20, 26.56){\circle*{0.5}}
 \put( 59.30, 26.54){\circle*{0.5}}
 \put( 59.40, 26.51){\circle*{0.5}}
 \put( 59.50, 26.49){\circle*{0.5}}
 \put( 59.60, 26.47){\circle*{0.5}}
 \put( 59.70, 26.45){\circle*{0.5}}
 \put( 59.80, 26.43){\circle*{0.5}}
 \put( 59.90, 26.40){\circle*{0.5}}
 \put( 60.00, 26.38){\circle*{0.5}}
 \put( 60.10, 26.36){\circle*{0.5}}
 \put( 60.20, 26.34){\circle*{0.5}}
 \put( 60.30, 26.32){\circle*{0.5}}
 \put( 60.40, 26.30){\circle*{0.5}}
 \put( 60.50, 26.27){\circle*{0.5}}
 \put( 60.60, 26.25){\circle*{0.5}}
 \put( 60.70, 26.23){\circle*{0.5}}
 \put( 60.80, 26.21){\circle*{0.5}}
 \put( 60.90, 26.19){\circle*{0.5}}
 \put( 61.00, 26.17){\circle*{0.5}}
 \put( 61.10, 26.15){\circle*{0.5}}
 \put( 61.20, 26.13){\circle*{0.5}}
 \put( 61.30, 26.11){\circle*{0.5}}
 \put( 61.40, 26.09){\circle*{0.5}}
 \put( 61.50, 26.07){\circle*{0.5}}
 \put( 61.60, 26.05){\circle*{0.5}}
 \put( 61.70, 26.03){\circle*{0.5}}
 \put( 61.80, 26.01){\circle*{0.5}}
 \put( 61.90, 25.98){\circle*{0.5}}
 \put( 62.00, 25.96){\circle*{0.5}}
 \put( 62.10, 25.94){\circle*{0.5}}
 \put( 62.20, 25.92){\circle*{0.5}}
 \put( 62.30, 25.90){\circle*{0.5}}
 \put( 62.40, 25.88){\circle*{0.5}}
 \put( 62.50, 25.86){\circle*{0.5}}
 \put( 62.60, 25.85){\circle*{0.5}}
 \put( 62.70, 25.83){\circle*{0.5}}
 \put( 62.80, 25.81){\circle*{0.5}}
 \put( 62.90, 25.79){\circle*{0.5}}
 \put( 63.00, 25.77){\circle*{0.5}}
 \put( 63.10, 25.75){\circle*{0.5}}
 \put( 63.20, 25.73){\circle*{0.5}}
 \put( 63.30, 25.71){\circle*{0.5}}
 \put( 63.40, 25.69){\circle*{0.5}}
 \put( 63.50, 25.67){\circle*{0.5}}
 \put( 63.60, 25.65){\circle*{0.5}}
 \put( 63.70, 25.63){\circle*{0.5}}
 \put( 63.80, 25.61){\circle*{0.5}}
 \put( 63.90, 25.59){\circle*{0.5}}
 \put( 64.00, 25.58){\circle*{0.5}}
 \put( 64.10, 25.56){\circle*{0.5}}
 \put( 64.20, 25.54){\circle*{0.5}}
 \put( 64.30, 25.52){\circle*{0.5}}
 \put( 64.40, 25.50){\circle*{0.5}}
 \put( 64.50, 25.48){\circle*{0.5}}
 \put( 64.60, 25.46){\circle*{0.5}}
 \put( 64.70, 25.45){\circle*{0.5}}
 \put( 64.80, 25.43){\circle*{0.5}}
 \put( 64.90, 25.41){\circle*{0.5}}
 \put( 65.00, 25.39){\circle*{0.5}}
 \put( 65.10, 25.37){\circle*{0.5}}
 \put( 65.20, 25.35){\circle*{0.5}}
 \put( 65.30, 25.34){\circle*{0.5}}
 \put( 65.40, 25.32){\circle*{0.5}}
 \put( 65.50, 25.30){\circle*{0.5}}
 \put( 65.60, 25.28){\circle*{0.5}}
 \put( 65.70, 25.26){\circle*{0.5}}
 \put( 65.80, 25.25){\circle*{0.5}}
 \put( 65.90, 25.23){\circle*{0.5}}
 \put( 66.00, 25.21){\circle*{0.5}}
 \put( 66.10, 25.19){\circle*{0.5}}
 \put( 66.20, 25.18){\circle*{0.5}}
 \put( 66.30, 25.16){\circle*{0.5}}
 \put( 66.40, 25.14){\circle*{0.5}}
 \put( 66.50, 25.12){\circle*{0.5}}
 \put( 66.60, 25.11){\circle*{0.5}}
 \put( 66.70, 25.09){\circle*{0.5}}
 \put( 66.80, 25.07){\circle*{0.5}}
 \put( 66.90, 25.05){\circle*{0.5}}
 \put( 67.00, 25.04){\circle*{0.5}}
 \put( 67.10, 25.02){\circle*{0.5}}
 \put( 67.20, 25.00){\circle*{0.5}}
 \put( 67.30, 24.99){\circle*{0.5}}
 \put( 67.40, 24.97){\circle*{0.5}}
 \put( 67.50, 24.95){\circle*{0.5}}
 \put( 67.60, 24.94){\circle*{0.5}}
 \put( 67.70, 24.92){\circle*{0.5}}
 \put( 67.80, 24.90){\circle*{0.5}}
 \put( 67.90, 24.89){\circle*{0.5}}
 \put( 68.00, 24.87){\circle*{0.5}}
 \put( 68.10, 24.85){\circle*{0.5}}
 \put( 68.20, 24.84){\circle*{0.5}}
 \put( 68.30, 24.82){\circle*{0.5}}
 \put( 68.40, 24.80){\circle*{0.5}}
 \put( 68.50, 24.79){\circle*{0.5}}
 \put( 68.60, 24.77){\circle*{0.5}}
 \put( 68.70, 24.75){\circle*{0.5}}
 \put( 68.80, 24.74){\circle*{0.5}}
 \put( 68.90, 24.72){\circle*{0.5}}
 \put( 69.00, 24.71){\circle*{0.5}}
 \put( 69.10, 24.69){\circle*{0.5}}
 \put( 69.20, 24.67){\circle*{0.5}}
 \put( 69.30, 24.66){\circle*{0.5}}
 \put( 69.40, 24.64){\circle*{0.5}}
 \put( 69.50, 24.63){\circle*{0.5}}
 \put( 69.60, 24.61){\circle*{0.5}}
 \put( 69.70, 24.59){\circle*{0.5}}
 \put( 69.80, 24.58){\circle*{0.5}}
 \put( 69.90, 24.56){\circle*{0.5}}
 \put( 70.00, 24.55){\circle*{0.5}}

\put(71.5,25){$\psi(nS)$}
\end{picture}
\end{center}
\caption{The dependence of leptonic constants for the nS-levels of
bottomonium and charmonium, $f_{n}$, as it is calculated in
the Quasilocal Sum Rules.}
\label{fh1}
\end{figure}
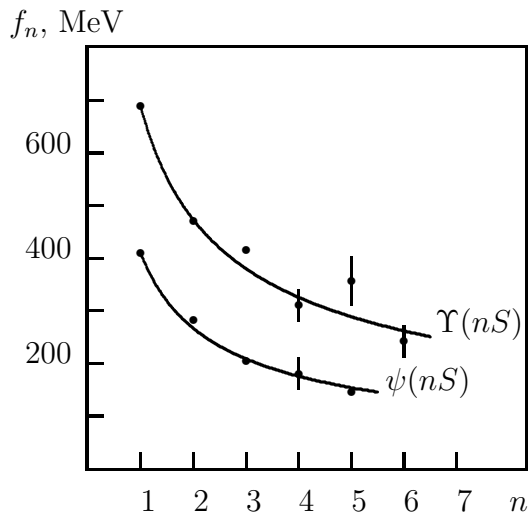

\begin{table}[bh]
\caption{The leptonic constants for the vector and pseudoscalar states of
$nS$-levels in the $(\bar b c)$-system, $f_n$ and $f_n^P$, calculated in the
sum rules resulting in the scaling relation. The accuracy is equal to 6\%.}
\label{t4}
\begin{center}
\begin{tabular}{||c|c|c||}
\hline
n & $f_n$, MeV  & $f_n^P$, MeV \\
\hline
1 & 385 & 397  \\
2 & 260 & 245 \\
\hline
\end{tabular}
\end{center}
\end{table}

\newpage
\centerline{\bf\sc 4. $\bf B_c$ decays}

\vspace*{5mm}
\underline{Lifetime.}\\
The $B_c$-meson decay processes can be subdivided into three classes:

1) the $\bar b$-quark decay with the spectator $c$-quark,

2) the $c$-quark decay 
with the spectator $\bar b$-quark and

3) the annihilation channel
$B_c^+\rightarrow l^+\nu_l (c\bar s, u\bar s)$, where $l=e,\; \mu,\; \tau$.

In the $\bar b \to \bar c c\bar s$ decays one separates 
also the \underline{Pauli interference}
with the $c$-quark from the initial state.
In accordance with the given classification, 
the total width is the sum over the partial widths
\begin{equation}
\Gamma (B_c\rightarrow X)=\Gamma (b\rightarrow X)
+\Gamma (c\rightarrow X)+\Gamma \mbox{(ann.)}+\Gamma\mbox{(PI)}\;.
\end{equation}
For the annihilation channel the  $\Gamma\mbox{(ann.)}$ width can
be reliably estimated in the framework of inclusive approach, where one
take the sum of the leptonic and quark decay modes with account for the hard
gluon corrections to the effective four-quark interaction of weak
currents. These corrections result in the factor of 
$a_1=1.22\pm 0.04$. The width is expressed through the
leptonic constant of $f_{B_c}=f_1^P\approx 400$ MeV. This estimate of the 
quark-contribution does not depend on a hadronization model, since the 
large energy release of the order of the meson mass takes place.
From the following expression one can see that one can neglect the contribution
by light leptons and quarks,
\begin{equation}
\Gamma \mbox{(ann.)} =\sum_{i=\tau,c}\frac{G^2_F}{8\pi}
|V_{bc}|^2f^2_{B_c}M m^2_i (1-m^2_i/m^2_{Bc})^2\cdot C_i\;,
\label{d3}
\end{equation}
where $C_\tau = 1$ for the $\tau^+\nu_\tau$-channel and 
$C_c =3|V_{cs}|^2a_1^2 $ for the $c\bar s$-channel.

As for the non-annihilation decays, in the approach of the {\sf operator 
product expansion} for the quark currents of weak decays \cite{ben}, 
one takes into account the $\alpha_s$-corrections to the free quark decays and
uses the quark-hadron duality for the final states. Then one considers the 
matrix element for the transition operator over the bound meson state.
The latter allows one also to
take into account effects caused by the motion and virtuality of decaying 
quark inside the meson because of the interaction with the spectator.
In this way the $\bar b\to \bar c c\bar s$ decay mode turns to be 
suppressed almost completely due to the Pauli interference with the charm quark
from the initial state. Besides, the $c$-quark decays with the 
spectator $\bar b$-quark is essentially suppressed in comparison with
the free quark decays because of a large bound energy in the initial state. 

In the framework of {\sf exclusive} approach it is necessary to sum widths 
of different decay modes calculated in the potential models \cite{sem,lus}. 
While considering the semileptonic decays due to the
$\bar b \to \bar c l^+\nu_l$ and $c\to s l^+\nu_l$ transitions
one finds that  in the former decays the hadronic final state is practically
saturated by the lightest bound $1S$-state in the $(\bar c c)$-system, 
i.e. by the $\eta_c$ and $J/\psi$ particles, and in the latter decays
the $1S$-states in the $(\bar b s)$-system, i.e. $B_s$ and 
$B_s^*$, only can enter the accessible energetic gap. 
The energy release in the latter transition is low in comparison with 
the meson masses, and, therefore, a visible deviation from the picture
of quark-hadron duality is possible. Numerical estimates show that the
value of $B_c\to(\bar b s)l^+\nu_l$ decay width is two times less 
in the exclusive approach than in the inclusive method, though this fact can
be caused by the choice of narrow wave package for the $B_s^{(*)}$ mesons in 
the quark model, so that $\tilde f_{B_s}\approx 150$ MeV, while
in the limit of static heavy quark, one should expect a larger value for the
leptonic constant. This increase will lead to the widening
of the wave package and, hence, to the increase of the overlapping integral
for the wave functions of $B_c$ and $B_s^{(*)}$.

Further, the $\bar b\to \bar c u\bar d$ channel, for example, can be calculated
through the given decay width of $\bar b \to \bar c l^+\nu_l$ with account
for the color factor and hard gluon corrections to the four-quark 
interaction. It can be also obtained as a sum over the widths of decays 
with the $(u\bar d)$-system bound states. 

\begin{table}[th]
\caption{The branching ratios of the $B_c$ decay modes calculated in
the framework of inclusive approach and in the exclusive quark model with 
the parameters $|V_{bc}|=0.040$, $\tilde f_{B_c}=0.47$ GeV, 
$\tilde f_{\psi}=0.54$
GeV, $\tilde f_{B_s}=0.3$ GeV, $m_b=4.8-4.9$ GeV, $m_c=1.5-1.6$ GeV,
$m_s=0.55$ GeV. The accuracy is about 10\%.}
\label{t5}
\begin{center}
\begin{tabular}{||l|c|c||}
\hline
$B_c$ decay mode & Inclus., \%  & Exclus., \%  \cite{bclhc}\\
\hline
$\bar b\to \bar c l^+\nu_l$ & 3.9 & 3.7 \\
$\bar b\to \bar c u\bar d$  & 16.2 & 16.7 \\
$\sum \bar b\to \bar c$     & 25.0 & 25.0 \\
$c\to s l^+\nu_l$           & 8.5  & 10.1 \\
$c\to s u\bar d$            & 47.3 & 45.4 \\
$\sum c\to s$               & 64.3 & 65.6 \\
$B_c^+\to \tau^+\nu_\tau$   & 2.9  & 2.0  \\
$B_c^+\to c\bar s$          & 7.2  & 7.2  \\
\hline
\end{tabular}
\end{center}
\end{table}

The results of calculation for the total $B_c$ width in the inclusive and
exclusive approaches give the values consistent with each other, if one 
takes into account the most significant uncertainty related with the choice
of quark masses (especially for the charm quark), so that finally we have 
\begin{equation}
\tau(B_c^+)= 0.55\pm 0.15\; \mbox{ps,}
\end{equation}
so that the observed $B_c$ candidates in the $\psi\pi$ mode at LEP
have quite a close value of the lifetime.

\underline{Exclusive decays.}\\
The consideration of exclusive $B_c$-decay modes supposes an introduction
of model for the hadronization of quarks into the mesons with the
given quantum numbers. The QCD sum rules for the three-point correlators of 
quark currents and potential models are 
among of those hadronization models.

A {\sf feature} of the sum rule application
to the mesons containing {\sf two heavy quarks}, is the account for the
significant 
role of the {\sf coulumb-like $\alpha_s/v$-corrections} due to the gluon
exchange 
between the quarks composing the meson and moving with the relative velocity
$v$. So, in the semileptonic decays of $B_c^+\to \psi(\eta_c)l^+\nu_l$, 
the heavy $(\bar Q_1Q_2)$ quarkonium is present in the both initial and 
final states, and, therefore, the contribution of coulumb-like corrections 
exhibits in a specially strong form. The use of tree approximation for 
the perturbative contribution into the three-point correlator of quark 
currents leads to a large deviation between the values of 
transition form-factors, calculated in the sum rules and 
potential models, respectively \cite{pav}. The account for the
$\alpha_s/v$-corrections removes this contradiction \cite{lep-sr}.

Thus, the meson potential models, based on the covariant expression for the
form-factors of weak $B_c$ decays through the {\sf overlapping} of quarkonium
wave 
functions in the initial and final states, and the QCD sum rules give 
the consistent description of semileptonic $B_c$-meson decays.  

Further, the {\sf hadronic} decay widths can be obtained on the basis of
assumption
on the factorization of the weak transition between the quarkonia and 
the hadronization of products of the virtual $W^{*+}$-boson decay \cite{fact}.
The accuracy of factorization has to raise with the increase of $W$-boson 
virtuality. This fact is caused by the suppression of interaction in the 
final state. In this way, the hadronic decays can be calculated due to the
use of form-factors for the semileptonic transitions with the 
relevant description of $W^*$ transition into the hadronic state.

\vspace*{7mm}
\begin{table}[bh]
\caption{The branching ratios of exclusive $B_c$ decay modes,
calculated in the framework of covariant quark model with the parameters 
$|V_{bc}|=0.040$, $\tilde f_{B_c}=0.47$ GeV, $\tilde f_{\psi}=0.54$ GeV,
$\tilde f_{B_s}=0.3$ GeV, $m_b=4.8-4.9$ GeV, $m_c=1.5-1.6$ GeV,
$m_s=0.55$ GeV. The accuracy equals 10\%.}
\label{t6}
\begin{center}
\begin{tabular}{||c|c|c|c||}
\hline
$B_c$ decay mode & BR, \%  & $B_c$ decay mode & BR, \% \\
\hline
$\psi l^+\nu_l$ & 2.5 & $\eta_c l^+\nu_l$ & 1.2 \\
$B^*_s l^+\nu_l$ & 6.2 & $B_s l^+\nu_l$    & 3.9 \\
$\psi \pi^+$    & 0.2 & $\eta_c \pi ^+$   & 0.2 \\
$B^*_s \pi^+$    & 5.2 & $B_s \pi^+$       & 5.5 \\
$\psi \rho^+$    & 0.6 & $\eta_c \rho ^+$   & 0.5 \\
$B^*_s \rho^+$    & 22.9 & $B_s \rho^+$       & 11.8 \\
\hline
\end{tabular}
\end{center}
\end{table}
More generally, the effective four-fermion hamiltonian for the nonleptonic
decays of
the $c$- and $b$-quarks has the form
\begin{eqnarray}
{\cal H}_{{\rm eff}}^c & = & \frac{G}{2\sqrt{2}}V_{uq_1}V_{cq_1}^*
[C_+^c(\mu)O_+^c+C_-^c(\mu)O_-^c]+h.c.\;, 
\label{ad 1} \\
{\cal H}_{{\rm eff}}^b & = & \frac{G}{2\sqrt{2}}V_{q_1b}V_{q_2q_3}^*
[C_+^b(\mu)O_+^b+C_-^b(\mu)O_-^b]+h.c. \;,
\label{ad 2}
\end{eqnarray}
where
$$
O^c_{\pm} = (\bar q_{1\alpha}\gamma_{\nu}(1-\gamma_5)c_\beta)
(\bar u_\gamma \gamma^\nu(1-\gamma_5)q_{2\delta})
(\delta_{\alpha\beta}\delta_{\gamma\delta}\pm\delta_{\alpha\delta}
\delta_{\gamma\beta}),
$$
$$
O^b_{\pm} = (\bar q_{1\alpha}\gamma_\nu(1-\gamma_5)b_\beta)
(\bar q_{3\gamma}\gamma^{\nu}(1-\gamma_5)q_{2\delta})
(\delta_{\alpha\beta}\delta_{\gamma\delta}\pm\delta_{\alpha\delta}
\delta_{\gamma\beta}).
$$

The factors $C_{\pm}^{c,b}(\mu)$ account for the strong corrections to the
corresponding four-fermion operators because of hard gluons.

In the leading logarithm approximation at $\mu > m_c$, one has
\begin{eqnarray}
C_+^c(\mu) & = &{\biggl(\frac{\alpha_s(M_W^2)}{\alpha_s(m_b^2)}\biggr)}^{6/23}
 {\biggl(\frac{\alpha_s(m_b^2)}{\alpha_s(\mu^2)}\biggr)}^{6/25}, \nonumber \\
C_-^c(\mu) & = & [C_+^c(\mu)]^{-2}. 
\label{ad 4}
\end{eqnarray}
At $\mu > m_b$, one finds
\begin{eqnarray}
C_+^b(\mu) & = &{\biggl(\frac{\alpha_s(M_W^2)}{\alpha_s(m_b^2)}\biggr)}^{6/23}
 {\biggl(\frac{\alpha_s(m_b^2)}{\alpha_s(\mu^2)}\biggr)}^{-3/25},  \\
C_-^b(\mu) & = &
{\biggl(\frac{\alpha_s(M_W^2)}{\alpha_s(m_b^2)}\biggr)}^{-12/23}
 {\biggl(\frac{\alpha_s(m_b^2)}{\alpha_s(\mu^2)}\biggr)}^{-12/25}.  
\end{eqnarray}

The $a_1$ and $a_2$ coefficients, accounting for the renormalization of the
four-fermion operators, are defined in the following way
\begin{eqnarray}
a_1 & = & C_+\frac{N_c+1}{2N_c}+C_-\frac{N_c-1}{2N_c},\\
a_2 & = & C_+\frac{N_c+1}{2N_c}-C_-\frac{N_c-1}{2N_c}.
\end{eqnarray}
In the limit $N_c\rightarrow \infty$ one has
\begin{eqnarray}
a_1 & \approx & 0.5\cdot(C_+ + C_-), \nonumber\\
a_2 & \approx & 0.5\cdot(C_+-C_-).
\end{eqnarray}
In the theory $a_2/a_1\sim 0.2$, which is in agreement with the experimental
data, so that the $B_c$ decay modes proportional to $a_2^2$ are
essentially suppressed.

\begin{figure}[th]
\vspace*{-45mm}
\hspace*{-5cm}
\epsfxsize=24cm \epsfbox{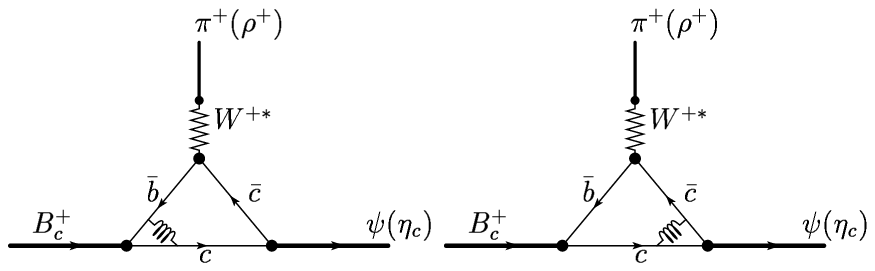}
\vspace*{-24.5cm}
\caption{The diagrams of $B_c^+\to \psi(\eta_c)\pi^+(\rho^+)$ decays
with account for the hard gluon exchange between the constituents.}
\label{hs}
\end{figure}

A decrease of the invariant mass for the hadron system  results in an
increase of the recoil meson momentum. This causes the problem of applicability
for the formalism of overlapping for the quarkonium wave functions, 
because, in this kinematics, the narrow wave packages are displaced relative 
to each other in the momentum space into the range of distribution tails.
In this situation one has to take into account a hard gluon exchange
between the quarkonium constituents, which destroys the spectator picture of 
weak transition in the potential approach.

The widths of
$B_c^+\to \psi \pi^+$ and $B_c^+\to \eta_c \pi^+$ decays \cite{hard} 
have the following forms
\begin{eqnarray}
\Gamma(B_c^+\to \psi \pi^+) &=& G_F^2 |V_{bc}|^2\; \frac{128\pi\alpha_s^2}{81}
 \biggl(\frac{M+m}{M-m}
\biggr)^3\; \frac{f^2_\pi \tilde f^2_{B_c}\tilde f^2_{\psi}M^3}
{(M-m)^2 m^2}\; a_1^2,
\\ 
\frac{\Gamma(B_c^+\to \eta_c \pi^+)}{\Gamma(B_c^+\to \psi \pi^+)} &=&
\frac{[5(M-m)^2(M+m)+(M-m)^3+8m^3]^2}
{16(M+m)^2M^4},
\label{g}
\end{eqnarray}
where $m=m_{\psi}$ or $m=m_{\eta_c}$, respectively.

Numerically one finds
\begin{eqnarray}
{\rm BR}^{\rm HS}(B_c^+\to \psi \pi^+) &=& 0.77\pm 0.19 \%, 
\label{br-}\\
{\rm BR}^{\rm HS}(B_c^+\to \eta_c \pi^+) &=& 1.00\pm 0.25 \%,\\
{\rm BR}^{\rm HS}(B_c^+\to \psi \rho^+) &=& 2.25\pm 0.56 \%, \\
{\rm BR}^{\rm HS}(B_c^+\to \eta_c \rho^+) &=& 2.78\pm 0.70 \%,
\label{br}
\end{eqnarray}
so that the matrix element is \underline{twice enhanced} in comparison with
that of
the potential model value due to the contribution of second
{\sf t}-channel diagram.

The relative yield of excited charmonium states can be also evaluated
\begin{equation}
\frac{{\rm BR}^{\rm HS}(B_c^+\to \psi(2S) \pi^+)}
{{\rm BR}^{\rm HS}(B_c^+\to \psi \pi^+)} =
\frac{{\rm BR}^{\rm HS}(B_c^+\to \eta_c(2S) \pi^+)}
{{\rm BR}^{\rm HS}(B_c^+\to \eta_c \pi^+) } \approx 0.36.
\end{equation}

As for the \underline{extraction of $B_c$ signal} in the hadronic background, 
the decay modes with $\psi$ in the final state are the most preferable, because
the latter particle can be easily identified by its leptonic decay mode.
This advantage is absent in the $B_c$ decay modes with the final state 
containing the $\eta_c$ or $B^{(*)}_s$ mesons, which are the objects, 
whose experimental registration is impeded by a large hadron background.

From the values of branching ratios shown above one can
easily obtain, that the total probability of $\psi$ yield in the $B_c$ decays
equals ${\rm BR}(B_c^+\to \psi X)=0.17$.

It is worth to note that the key
role in the $B_c$ signal observation plays the presence of the {\sf vertex
detector,}
which allows one to extract events with the weak decays of
long-lived particles containing heavy quarks. In the case under consideration
it gives the possibility to suppress the background from the direct $\psi$ 
production. In the semileptonic $B_c^+\to \psi l^+\nu_l$ decays 
the presence of vertex detector and large statistics of events allows one
to determine the $B_c$-meson mass and to separate the events with its decays 
from the ordinary $B_{u}^+$-meson decays, which have no $\psi l^+$ mode. 
In the $\psi\pi^+$ decay the direct measurement of $B_c$ mass is possible.
The detector efficiency in the reconstruction of three-particle secondary
vertex ($l^+l^-$  from decays of $\psi$ and $\pi^+$ or $l^+$) becomes the
most important characteristics here. The low efficiency of the LEP detectors 
($\epsilon\approx 0.15$), for example, makes the $B_c$ observation to be
hardly reachable in the experiments at the electron-positron collider.

\vspace*{3mm}
\centerline{\bf\sc 5. $\bf B_c$ production}

\vspace*{2.6mm}
The $(\bar b c)$ system is a heavy quarkonium, i.e. it contains two heavy
quarks.
This determines the general features for the $B_c$ meson production in various
interactions:

{\sf 1.} Perturbative calculations for the hard associative production of
two heavy pairs of $\bar c c$ and $\bar b b$ cause the suppressed $B_c$ yield
as a $10^{-3}$ fraction of beauty hadrons.

{\sf 2.} A soft nonperturbative binding of nonrelativistic quarks in the
color-singlet
state can be described in the framework of potential models.

The latter means the factorization of hard and soft amplitudes, so that the
soft factor
is determined by the radial wave function at the origin for the S-wave
quarkonium and its
first derivative for the P-wave one. Thus, the analysis of $B_c$ production
mechanisms is reduced to the consideration of perturbative amplitudes in
high orders over $\alpha_s$.
\vspace*{3mm}
\newpage
\centerline{\underline{$e^+e^-$ annihilation}}

\vspace*{3mm}
In $e^+e^-$-annihilation at large energies ($M^2_{B_c}/s\ll 1$),
the consideration of leading diagrams for the $B_c$-meson production gives 
the factorized scaling result for the differential cross-section over  
the energy fraction carried out by meson,
$d\sigma/dz= \sigma(\bar b b)\cdot D(z)$, where $z=2E_{B_c}/\sqrt{s}$.
This distribution allows the interpretation as
the hard production of heavier $\bar b$-quark with the subsequent fragmentation
into $B_c$, so that $D(z)$ is the fragmentation function. In this approach
one can obtain analytic expressions for the fragmentation functions into
the different spin states of $S$, $P$, and $D$-wave levels \cite{frag}.

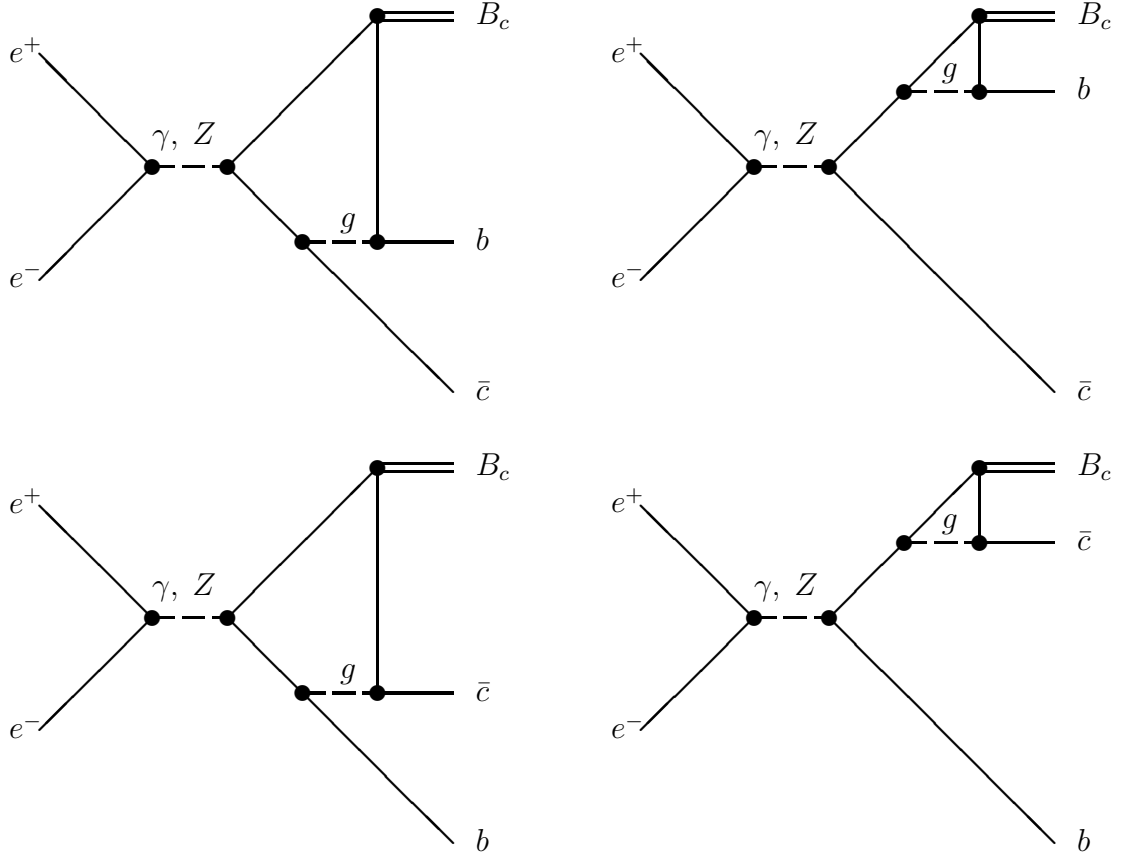
\begin{figure}[th]
\setlength{\unitlength}{1mm}\thicklines

\begin{center}
\begin{picture}(135,120)
\put(20,100){\circle*{2}}
\put(20,100){\line(-1,1){15}}
\put(20,100){\line(-1,-1){15}}
\put(1,114){$e^+$}
\put(1,84){$e^-$}
\put(20,100){\line(1,0){3}}
\put(24,100){\line(1,0){3}}
\put(28,100){\line(1,0){2}}
\put(30,100){\circle*{2}}
\put(30,100){\line(1,1){20}}
\put(30,100){\line(1,-1){30}}
\put(50,120){\line(0,-1){30}}
\put(50,90){\line(1,0){10}}
\put(50,120.5){\line(1,0){10}}
\put(50,119.5){\line(1,0){10}}
\put(40,90){\line(1,0){3}}
\put(44,90){\line(1,0){3}}
\put(48,90){\line(1,0){2}}
\put(50,120){\circle*{2}}
\put(50,90){\circle*{2}}
\put(40,90){\circle*{2}}
\put(63,119){$B_c$}
\put(63,89){$b$}
\put(63,69){$\bar c$}
\put(20,103){$\gamma,\; Z$}
\put(45,92){$g$}

\put(100,100){\circle*{2}}
\put(100,100){\line(-1,1){15}}
\put(100,100){\line(-1,-1){15}}
\put(81,114){$e^+$}
\put(81,84){$e^-$}
\put(100,100){\line(1,0){3}}
\put(104,100){\line(1,0){3}}
\put(108,100){\line(1,0){2}}
\put(110,100){\circle*{2}}
\put(110,100){\line(1,1){20}}
\put(110,100){\line(1,-1){30}}
\put(130,120){\line(0,-1){10}}
\put(130,110){\line(1,0){10}}
\put(130,120.5){\line(1,0){10}}
\put(130,119.5){\line(1,0){10}}
\put(120,110){\line(1,0){3}}
\put(124,110){\line(1,0){3}}
\put(128,110){\line(1,0){2}}
\put(130,120){\circle*{2}}
\put(130,110){\circle*{2}}
\put(120,110){\circle*{2}}
\put(143,119){$B_c$}
\put(143,109){$b$}
\put(143,69){$\bar c$}
\put(100,103){$\gamma,\; Z$}
\put(125,112){$g$}

\put(20,40){\circle*{2}}
\put(20,40){\line(-1,1){15}}
\put(20,40){\line(-1,-1){15}}
\put(1,54){$e^+$}
\put(1,24){$e^-$}
\put(20,40){\line(1,0){3}}
\put(24,40){\line(1,0){3}}
\put(28,40){\line(1,0){2}}
\put(30,40){\circle*{2}}
\put(30,40){\line(1,1){20}}
\put(30,40){\line(1,-1){30}}
\put(50,60){\line(0,-1){30}}
\put(50,30){\line(1,0){10}}
\put(50,60.5){\line(1,0){10}}
\put(50,59.5){\line(1,0){10}}
\put(40,30){\line(1,0){3}}
\put(44,30){\line(1,0){3}}
\put(48,30){\line(1,0){2}}
\put(50,60){\circle*{2}}
\put(50,30){\circle*{2}}
\put(40,30){\circle*{2}}
\put(63,59){$B_c$}
\put(63,29){$\bar c$}
\put(63,9){$b$}
\put(20,43){$\gamma,\; Z$}
\put(45,32){$g$}

\put(100,40){\circle*{2}}
\put(100,40){\line(-1,1){15}}
\put(100,40){\line(-1,-1){15}}
\put(81,54){$e^+$}
\put(81,24){$e^-$}
\put(100,40){\line(1,0){3}}
\put(104,40){\line(1,0){3}}
\put(108,40){\line(1,0){2}}
\put(110,40){\circle*{2}}
\put(110,40){\line(1,1){20}}
\put(110,40){\line(1,-1){30}}
\put(130,60){\line(0,-1){10}}
\put(130,50){\line(1,0){10}}
\put(130,60.5){\line(1,0){10}}
\put(130,59.5){\line(1,0){10}}
\put(120,50){\line(1,0){3}}
\put(124,50){\line(1,0){3}}
\put(128,50){\line(1,0){2}}
\put(130,60){\circle*{2}}
\put(130,50){\circle*{2}}
\put(120,50){\circle*{2}}
\put(143,59){$B_c$}
\put(143,49){$\bar c$}
\put(143,9){$b$}
\put(100,43){$\gamma,\; Z$}
\put(125,52){$g$}

\end{picture}
\end{center}
\vspace*{-12mm}
\caption{The diagrams of the $B_c$-meson production in
$e^+e^-$-annihilation.}
\label{fp1}
\end{figure}
\noindent Denoting $r=m_c/(m_b+m_c)$, for the pseudoscalar state one finds
\begin{eqnarray}
D(z)_{\bar b\rightarrow B_c} & = &
\frac{8\alpha^2_s |\Psi (0)|^2}{81 m^3_c}
\frac{rz(1-z)^2}{(1-(1-r)z)^6}
(6-18(1-2r)z+(21-74r+68r^2)z^2 - \nonumber \\
&&
2(1-r)(6-19r+18r^2)z^3 +3(1-r)^2 (1-2r+2r^2)z^4),\label{p6}
\end{eqnarray}
for the vector meson the perturbative fragmentation function is equal to
\begin{eqnarray}
D(z)_{\bar b\rightarrow B^*_c} & = &
\frac{8\alpha^2_s |\Psi (0)|^2 }{27 m^3_c}
\frac{rz(1-z)^2}{(1-(1-r)z)^6}
(2-2(3-2r)z+ 3(3-2r +4r^2)z^2 - \nonumber \\
&&
2(1-r)(4-r+2r^2)z^3 + (1-r)^2(3-2r+2r^2)z^4),\label{p7}
\end{eqnarray}

These expressions take into account the spin structure of both
interactions and bound states, but they are very close to the Peterson et al.
model \cite{peter}, where
$$
D(z)_{\bar b\rightarrow B^{(*)}_c} \sim \frac{1}{z}
\frac{1}{(m_b^2-\frac{M^2}{z}-\frac{m_c^2}{1-z})^2}\sim
\frac{rz(1-z)^2}{(1-(1-r)z)^4},
$$
so that the calculated functions are slightly more hard.
Numerically, one finds $\tilde f(\bar b \to B_c^+) = \sigma(B_c^+)/\sigma(b\bar
b) = (1.3\pm 0.4)\cdot 10^{-3}$.
Combining the latter with the prediction of branching ratios one gets
$$
[f(\bar b \to B_c^+)\cdot {\rm BR}(B_c^+\to \psi \pi^+)]_{\rm TH} =
(0.22\pm 0.09)\cdot 10^{-5},
$$
where $f(\bar b \to B_c^+) = \sigma(Z\to B_c^+)/\sigma(Z\to q\bar q)$,
which may be compared with the OPAL estimate
$$
[f(\bar b \to B_c^+)\cdot {\rm BR}(B_c^+\to \psi \pi^+)]_{\rm OPAL} =
(3.8^{+5.0}_{-2.4}\pm 0.5)\cdot 10^{-5}.
$$
If one takes into account both the single DELPHI candidate in the $\psi\pi$
mode and
absence of the event at ALEPH, then assuming the equal detector-efficiencies,
one
estimates
$$
[f(\bar b \to B_c^+)\cdot {\rm BR}(B_c^+\to \psi \pi^+)]_{\rm LEP} =
(1.9^{+2.5}_{-1.2}\pm 0.3)\cdot 10^{-5}.
$$

\begin{figure}[ph]
\hspace*{1cm}$d\sigma_{e^+e^-}/dz$, pb\\
\vspace*{-9mm}
\begin{center}
\hspace*{-10mm}
\epsfxsize=15cm \epsfbox{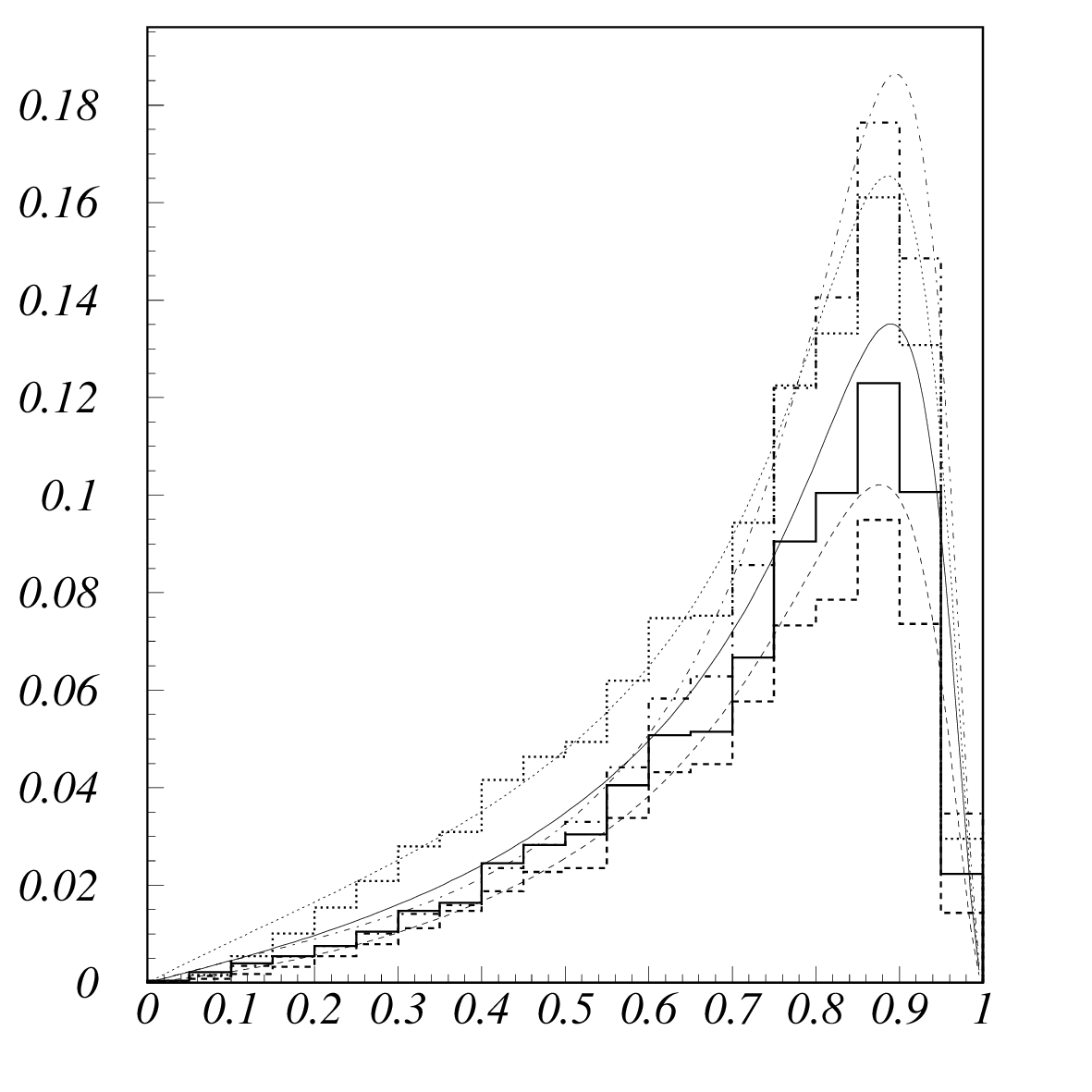}
\end{center}
\vspace*{-13mm}
{\hfill $z$\hspace*{8mm}}
\caption{The differential cross-sections over $z$ for the $B_c$ production
in  $e^+e^- \rightarrow \gamma^* \rightarrow  B_c+X$ (histograms)
in comparison with the fragmentation contribution calculated analytically
(smooth curves) at 100 GeV for the following states of $B_c$: $^1P_1$(solid
line),
$^3P_0$(dashed line), $^3P_1$(dotted line),
$^3P_2$(dash-dotted line).}
\label{g3fp2}
\end{figure}

\centerline{\underline{Photon-photon collisions}}

\vspace*{3mm}
In photon-photon interactions for the $B_c$ production in the leading
approximation of perturbation theory, one can isolate three
gauge-invariant groups of diagrams, which can be interpreted as 

{\bf 1.} the hard photon-photon production of $b\bar b$ with 
the subsequent fragmentation of $\bar b \to B_c^+(nL)$, 
where $n$ is the principal quantum number of the $(\bar b c)$-quarkonium, 
$L=0,1\ldots$ is the orbital angular momentum, 

{\bf 2.} the corresponding production and fragmentation for the $c$-quarks, and 

{\bf 3.} the recombination diagrams of $(\bar b c)$-pair
into $B_c^+$, wherein the quarks of different flavours are 
connected to the different photon lines.

In this case, the results of calculation for the complete set of diagrams
in the leading order of perturbation theory show that the group of
$b$-fragmentation diagrams at high transverse momenta
$p_T(B_c)\gg M_{B_c}$ can be described by the fragmentation model with the
fragmentation function $D_{\bar b\to B_c^+}(z)$, calculated in the 
$e^+e^-$-annihilation. The set of $c$-fragmentation diagrams does not
allow the description in the framework of fragmentation model. The 
recombination diagrams give the dominant contribution to the total 
cross-section for the photon-photon production of $B_c$ \cite{phot}.

The reason is that the photon coupled to the charm quark can be considered
as the "resolved" one, so that the bottom quark is hardly produced off the
charm quark appearing in the structure function of the photon, as it is
evaluated in the lowest order of perturbative theory.

\begin{figure}[ph]
\epsfxsize=17cm \epsfbox{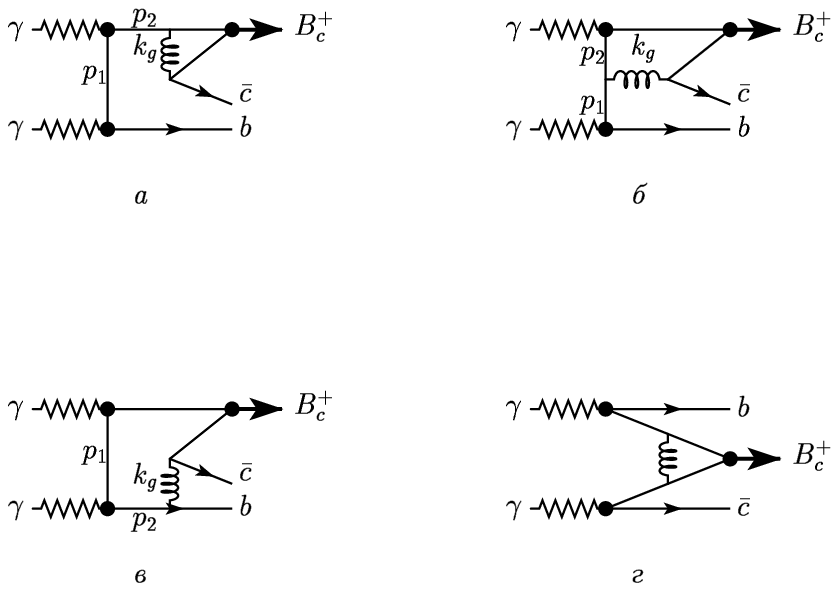}
\vspace*{-85mm}
\caption{The contribution of fragmentation for $\bar b\to B_c^+$ in the
photonic
production (three initial diagrams) and the recombination (the last diagram).}
\label{g3tf1}
\end{figure}

\begin{figure}[p]
\hspace*{1cm}$\sigma$, fb\\
\vspace*{-9mm}
\begin{center}
\hspace*{-10mm}
\epsfxsize=15cm \epsfbox{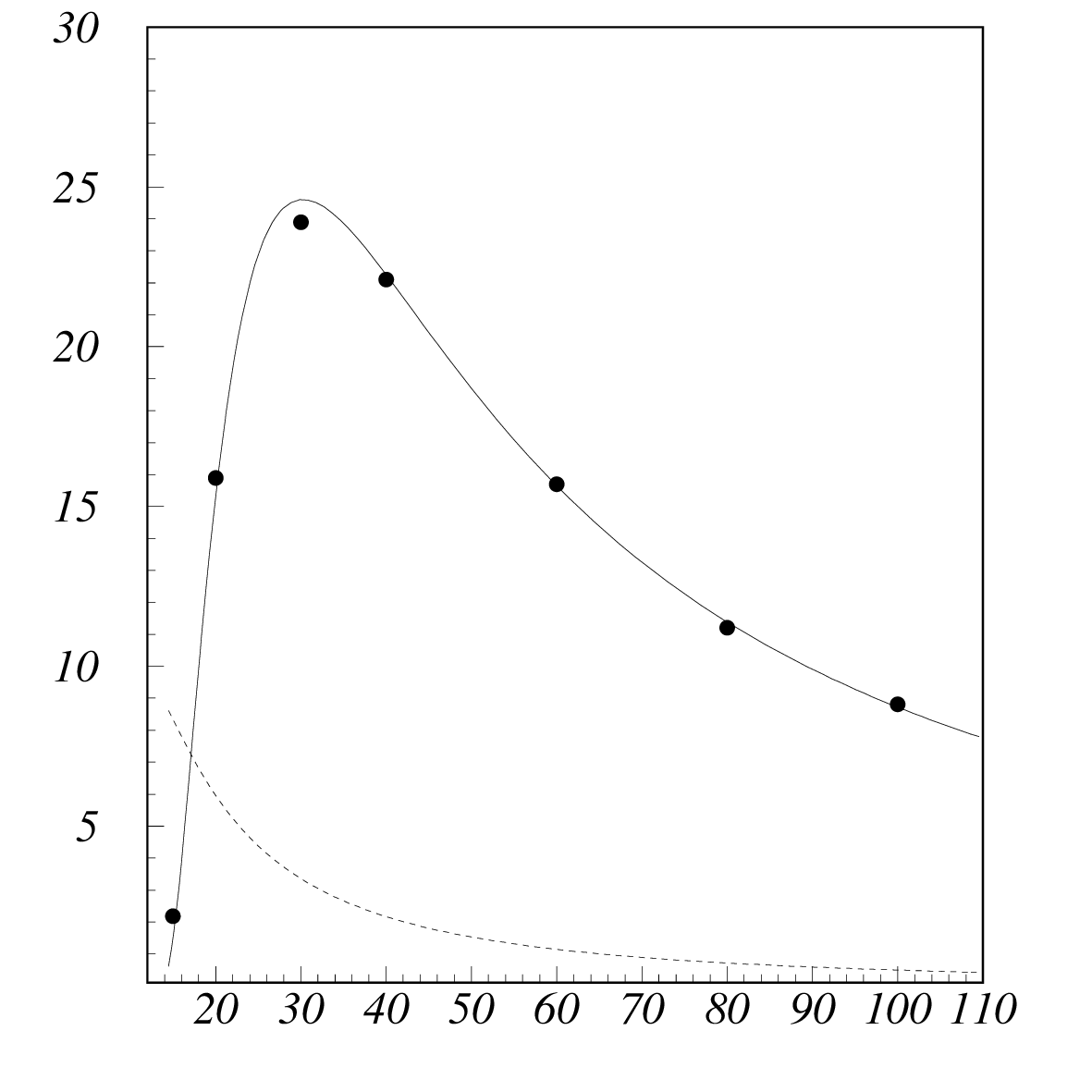}
\end{center}
\vspace*{-13mm}
{\hfill GeV\hspace*{8mm}}
\caption{The summed total cross-section versus the energy of photonic
production for the P-wave states of $B_c$
$(\bullet)$ in comparison with the fragmentation contribution
(dashed line). The solid-line curve is the fit of points.}
\label{g3tf3}
\end{figure}

\begin{figure}[p]
\vspace*{-1.5cm}
\hspace*{-0.7cm}\parbox{16cm}{%
\parbox{16cm}{%
\parbox[t]{7.7cm}{%
\hbox to 1.5cm {\hfil\mbox{$d\sigma_{\gamma\gamma}/dz$, fb}}
\epsfxsize=7.5cm \epsfbox{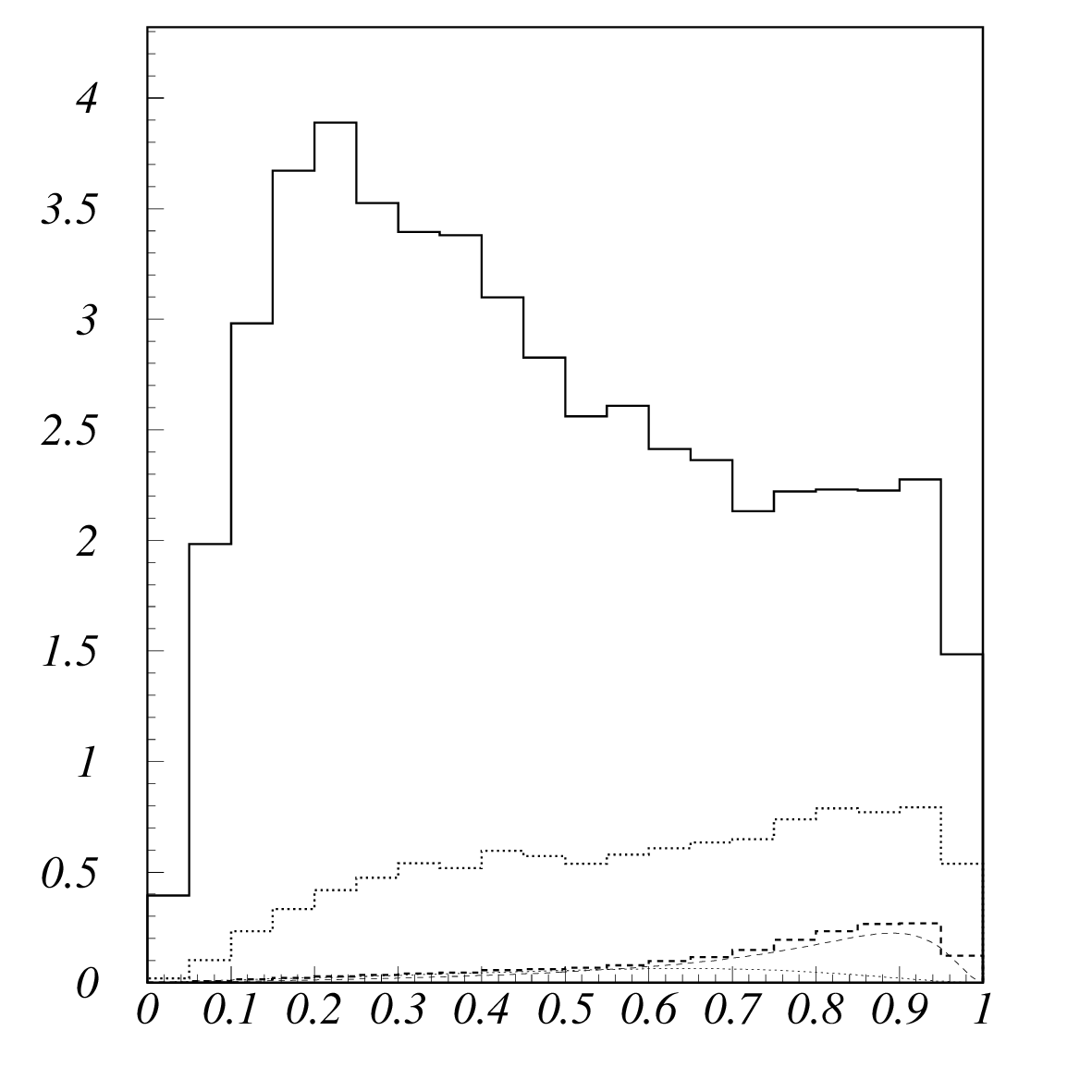}

\vspace*{-0.6cm}
\hbox to 7.0cm {\hfil\mbox{$z$}}
\vspace*{-0.6cm}
\parbox{7.7cm}{\bf \hfil $^1P_1$ \hfil}
}
\parbox{0.6cm}{ }
\parbox[t]{7.7cm}{%
\hbox to 1.5cm {\hfil\mbox{$d\sigma_{\gamma\gamma}/dz$, fb}}
\epsfxsize=7.5cm \epsfbox{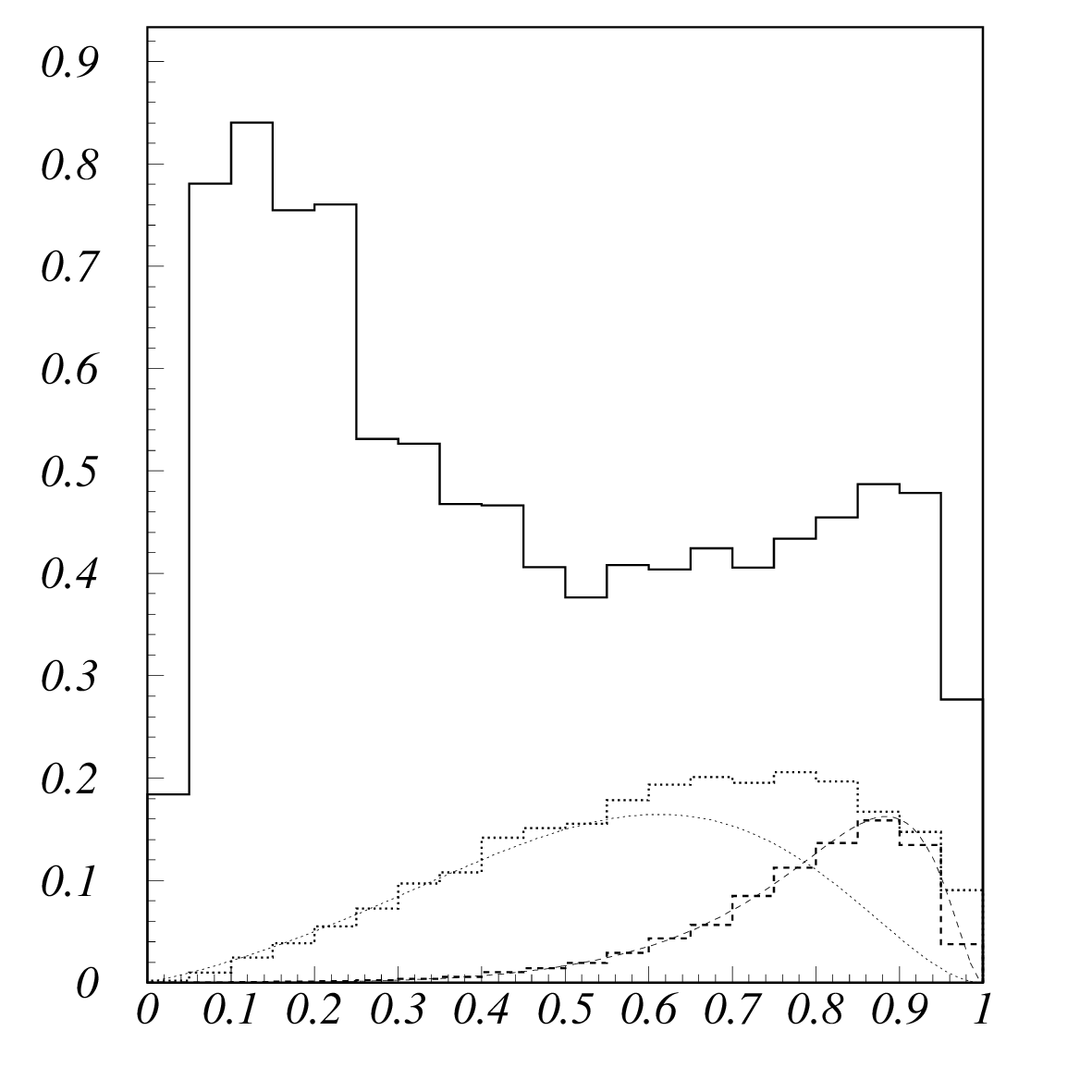}

\vspace*{-0.6cm}
\hbox to 7.0cm {\hfil\mbox{$z$}}
\vspace*{-0.6cm}
\parbox{7.7cm}{\bf \hfil $^3P_0$ \hfil}
}}

\vspace{0.7cm}
\parbox{16cm}{%
\parbox[t]{7.7cm}{%
\hbox to 1.5cm {\hfil\mbox{$d\sigma_{\gamma\gamma}/dz$, fb}}
\epsfxsize=7.5cm \epsfbox{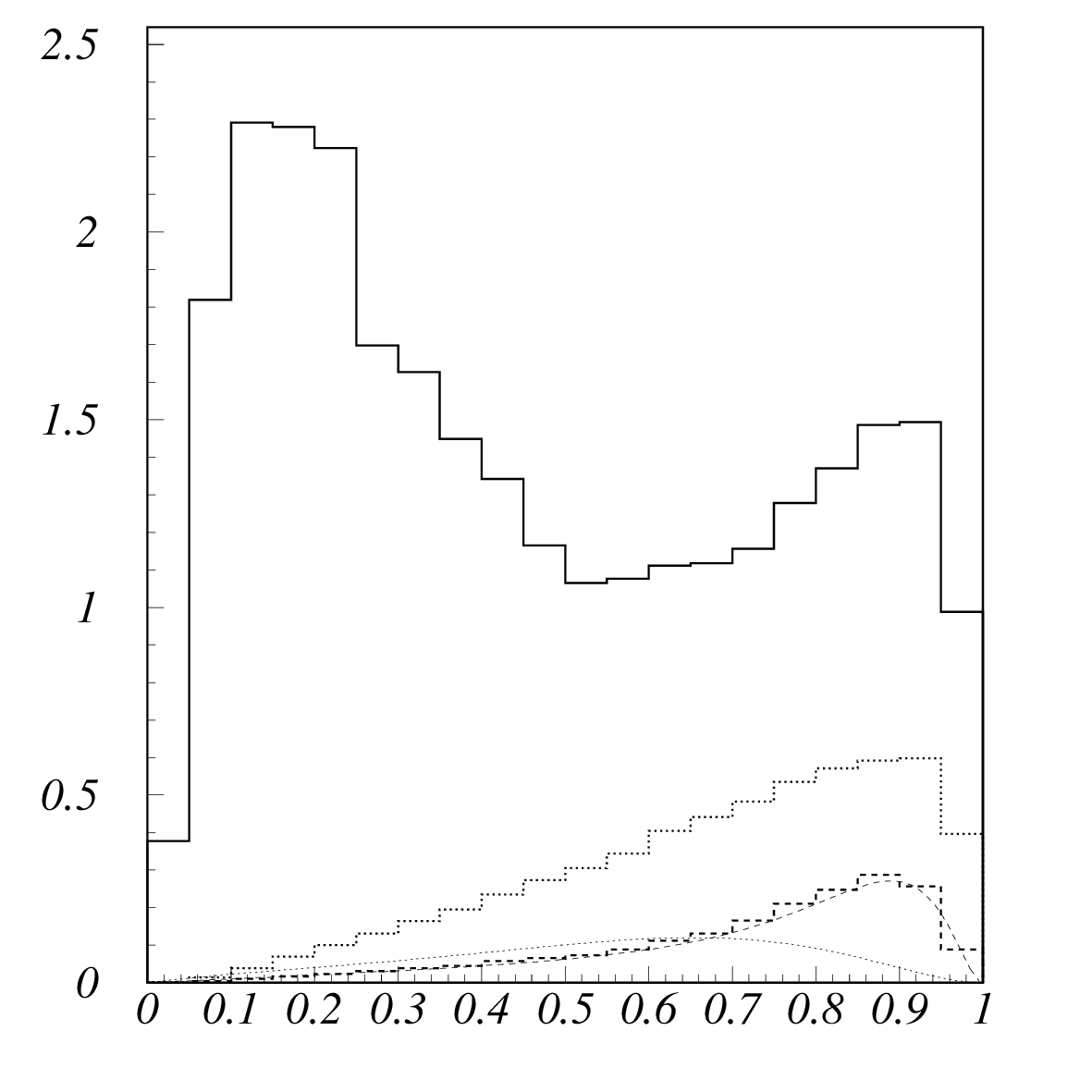}

\vspace*{-0.6cm}
\hbox to 7.0cm {\hfil\mbox{$z$}}
\vspace*{-0.6cm}
\parbox{7.7cm}{\bf \hfil $^3P_1$ \hfil}
}
\parbox{0.6cm}{ }
\parbox[t]{7.7cm}{%
\hbox to 1.5cm {\hfil\mbox{$d\sigma_{\gamma\gamma}/dz$, fb}}

\epsfxsize=7.5cm \epsfbox{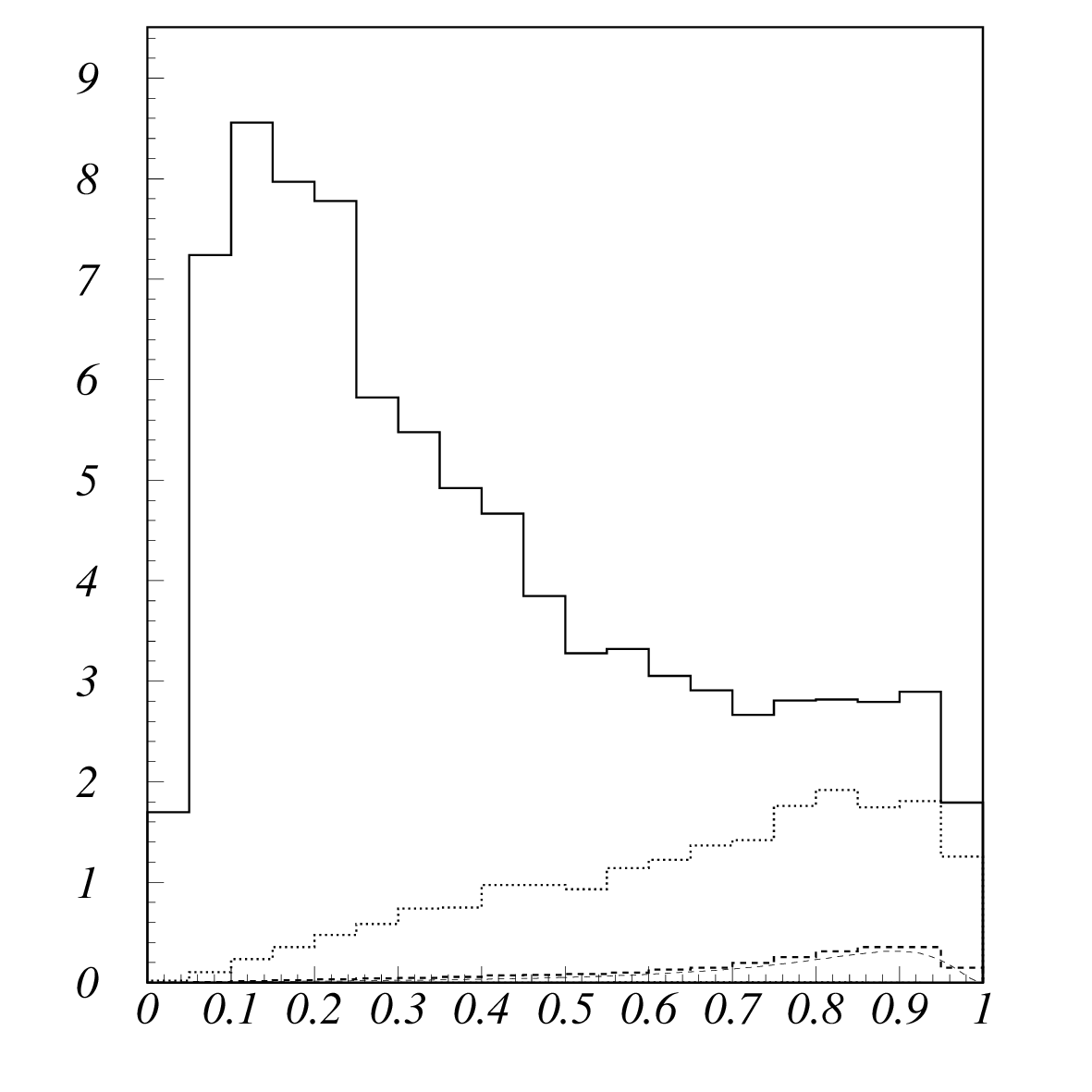}

\vspace*{-0.6cm}
\hbox to 7.0cm {\hfil\mbox{$z$}}
\vspace*{-0.6cm}
\parbox{7.7cm}{\bf \hfil $^3P_2$ \hfil}
}}

\vspace{-2mm}
\caption{The differential distributions over $z$ for the photonic production
of P-wave states of $B_c$ at 100 GeV:
complete set of diagrams (solid histogram),
the $\bar b$-fragmentation diagrams in comparison with the
prediction of fragmentation model (dashed histogram and curve),
the $c$-fragmentation diagrams and the fragmentation model (dotted histogram
and curve), correspondingly.}
\label{g3tf4}}
\end{figure}

\begin{figure}[p]
\vspace*{1.5cm}
\hspace*{-0.7cm}
\parbox{16cm}{%
\parbox{16cm}{%
\parbox[t]{7.7cm}{%
\hbox to 1.5cm {\hfil\mbox{$d\sigma_{\gamma\gamma}/dp_T$, fb/GeV}}
\epsfxsize=7.5cm \epsfbox{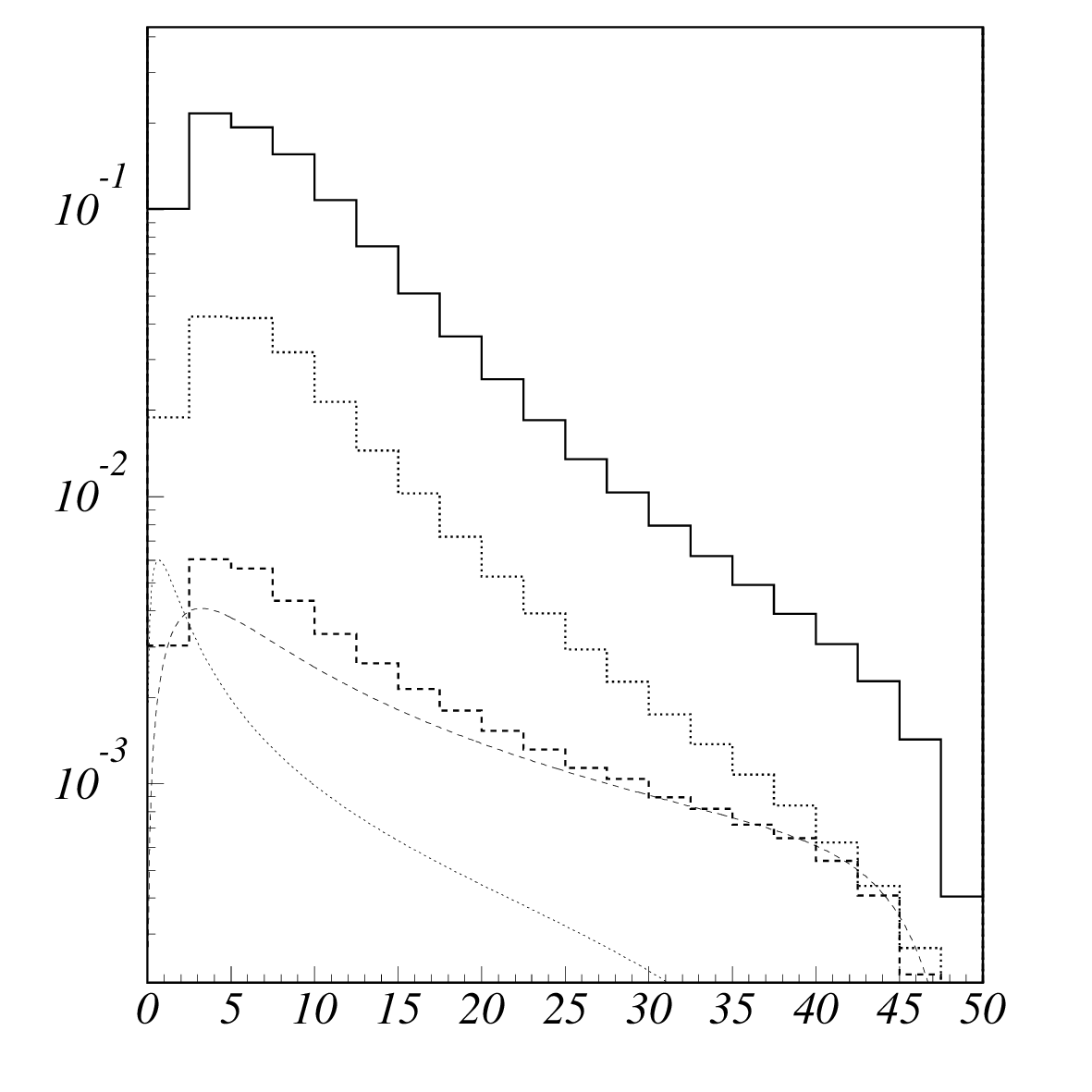}

\vspace*{-0.6cm}
\hbox to 7.7cm {\hfil\mbox{$p_T$, GeV}}
\vspace*{-0.6cm}
\parbox{7.7cm}{\bf \hfil $^1P_1$ \hfil}
}
\parbox{0.6cm}{ }
\parbox[t]{7.7cm}{%
\hbox to 1.5cm {\hfil\mbox{$d\sigma_{\gamma\gamma}/dp_T$, fb/GeV}}
\epsfxsize=7.5cm \epsfbox{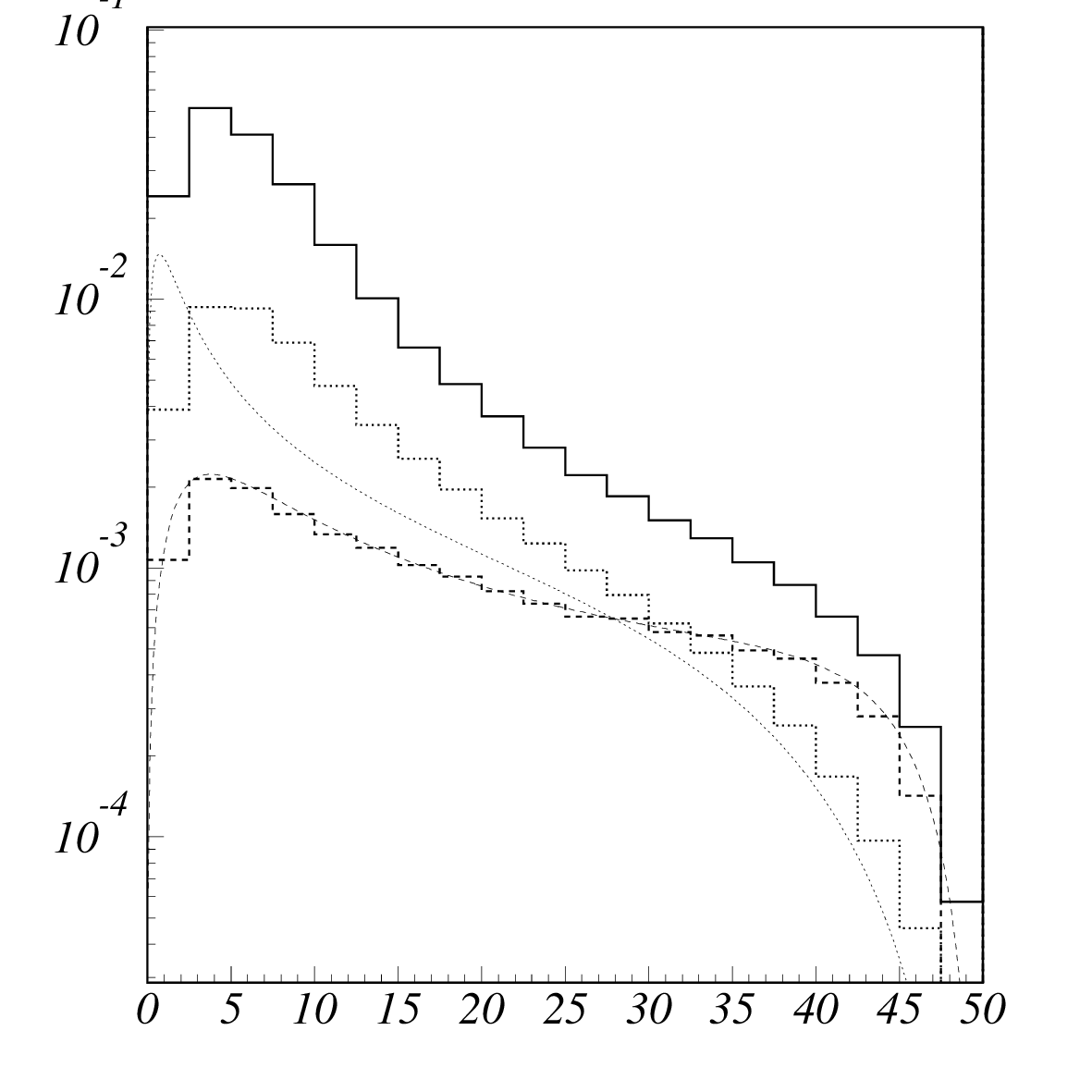}

\vspace*{-0.6cm}
\hbox to 7.7cm {\hfil\mbox{$p_T$, GeV}}
\vspace*{-0.6cm}
\parbox{7.7cm}{\bf \hfil $^3P_0$ \hfil}
}}

\vspace{0.5cm}
\parbox{16cm}{%
\parbox[t]{7.7cm}{%
\hbox to 1.5cm {\hfil\mbox{$d\sigma_{\gamma\gamma}/dp_T$, fb/GeV}}
\epsfxsize=7.5cm \epsfbox{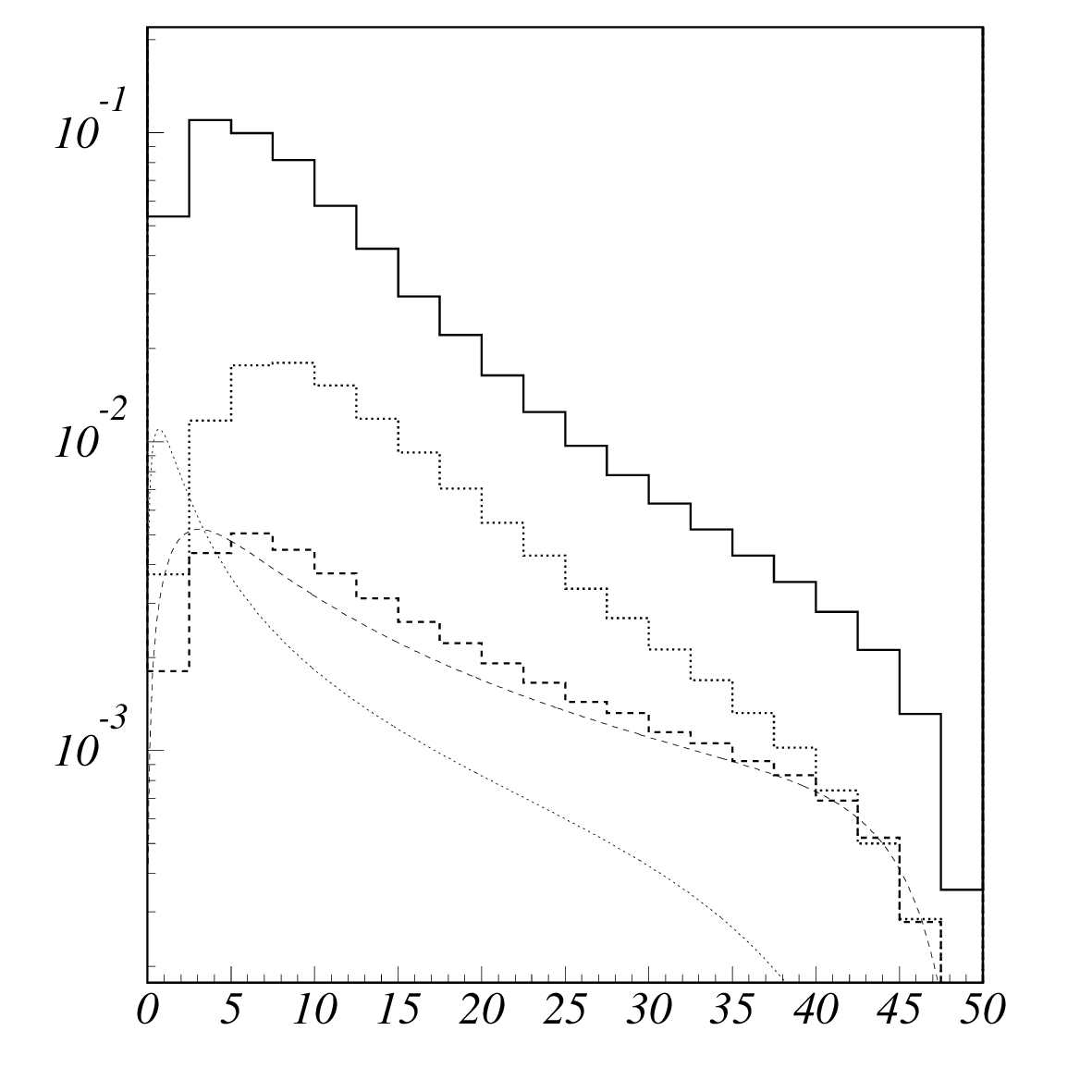}

\vspace*{-0.6cm}
\hbox to 7.7cm {\hfil\mbox{$p_T$, GeV}}
\vspace*{-0.6cm}
\parbox{7.7cm}{\bf \hfil $^3P_1$ \hfil}
}
\parbox{0.6cm}{ }
\parbox[t]{7.7cm}{%
\hbox to 1.5cm {\hfil\mbox{$d\sigma_{\gamma\gamma}/dp_T$, fb/GeV}}
\epsfxsize=7.5cm \epsfbox{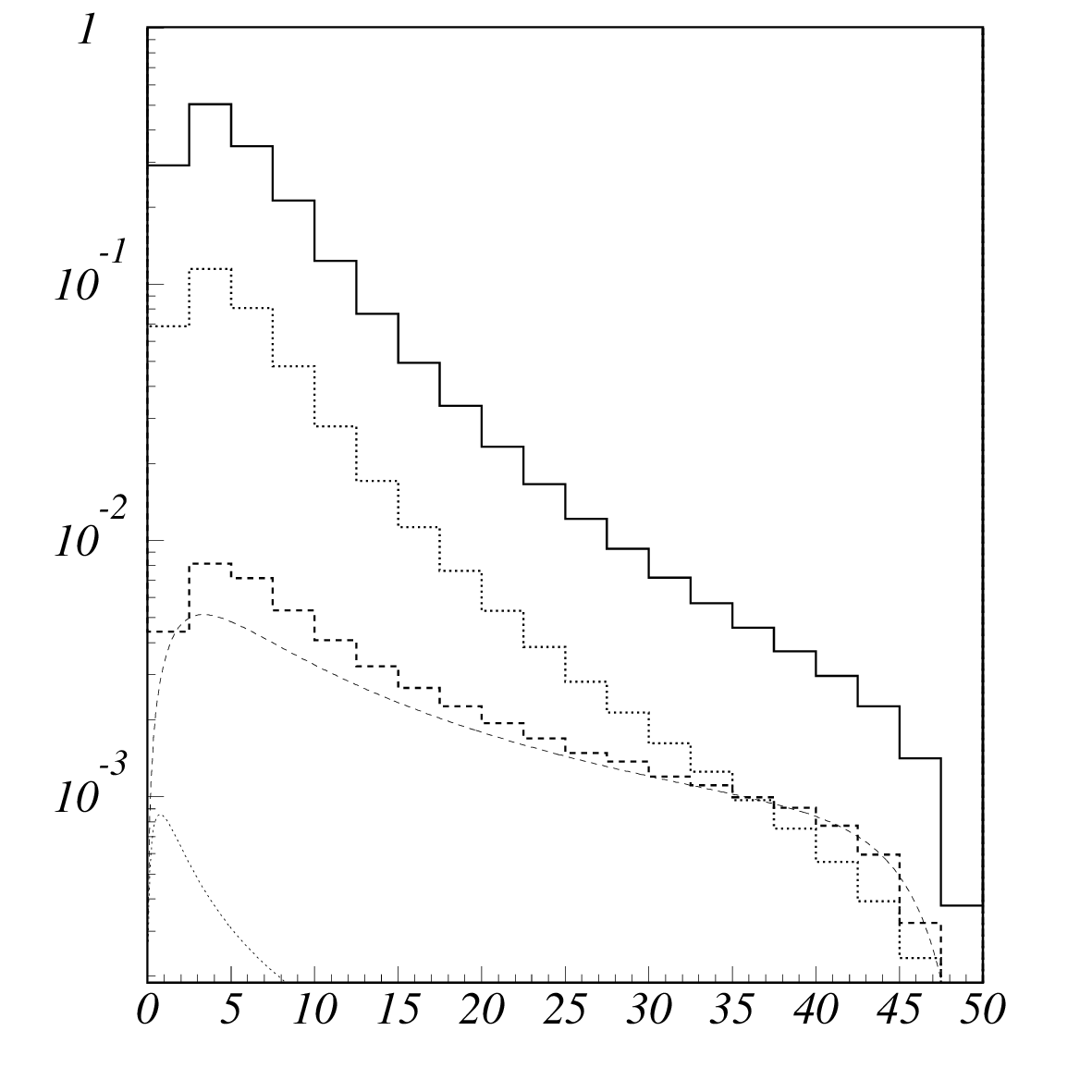}

\vspace*{-0.6cm}
\hbox to 7.7cm {\hfil\mbox{$p_T$, GeV}}
\vspace*{-0.6cm}
\parbox{7.7cm}{\bf \hfil $^3P_2$ \hfil}
}}

\caption{The distributions over the transverse momentum.}
\label{g3tf5}}
\end{figure}

\newpage
\centerline{\underline{Hadronic production of $B_c$}}

\vspace*{3mm}
The parton subprocess of gluon-gluon fusion 
$gg\to B_c^+ +b +\bar c$ dominates in the hadron-hadron production of 
$B_c$ mesons. In the leading approximation of QCD perturbation theory it 
requires  the calculation of 36 diagrams in the fourth order over the 
$\alpha_s$ coupling constant. In this case, there are no
isolated gauge-invariant groups of diagrams, which would allow the 
interpretation similar to the consideration of $B_c$ production in 
$e^+e^-$-annihilation and photon-photon collisions. 

By the general theorem on factorization,  it is clear that
at high transverse momenta the fragmentation of the heavier quark
$Q\to (Q\bar q) + q$ must dominate. It is described by the factorized 
formula
\begin{equation}
\frac{d\sigma}{dp_T} = \int \frac{d\hat \sigma(\mu; gg\to Q\bar Q)}
{dk_T}|_{k_T=p_T/x}\cdot D^{Q\to (Q\bar q)}(x;\mu)\; 
\frac{dx}{x}\;,
\label{one}
\end{equation}
where $\mu$ is the factorization scale, 
$d\hat\sigma/dk_T$ is the cross-section
for the gluon-gluon production of quarks $Q+\bar Q$, $D$ is the fragmentation
function. 

The calculation for the complete set of diagrams of the
$O(\alpha_s^4)$-contribution \cite{hadr}
allows one to determine a value of the transverse momentum $p_T^{min}$, 
being the low boundary of the region, where
the subprocess of gluon-gluon $B_c$-meson production enters 
the regime of factorization for the hard production of $b\bar b$-pair and 
the subsequent fragmentation of $\bar b$-quark into the bound 
$(\bar b c)$-state, as it follows from the theorem on the
factorization of the hard processes in the perturbative QCD. 

\begin{figure}[p]
\hspace*{1cm}$d\hat\sigma/dp_T$, nb/GeV\\
\vspace*{-5mm}
\begin{center}
\hspace*{-10mm}
\epsfxsize=14cm \epsfbox{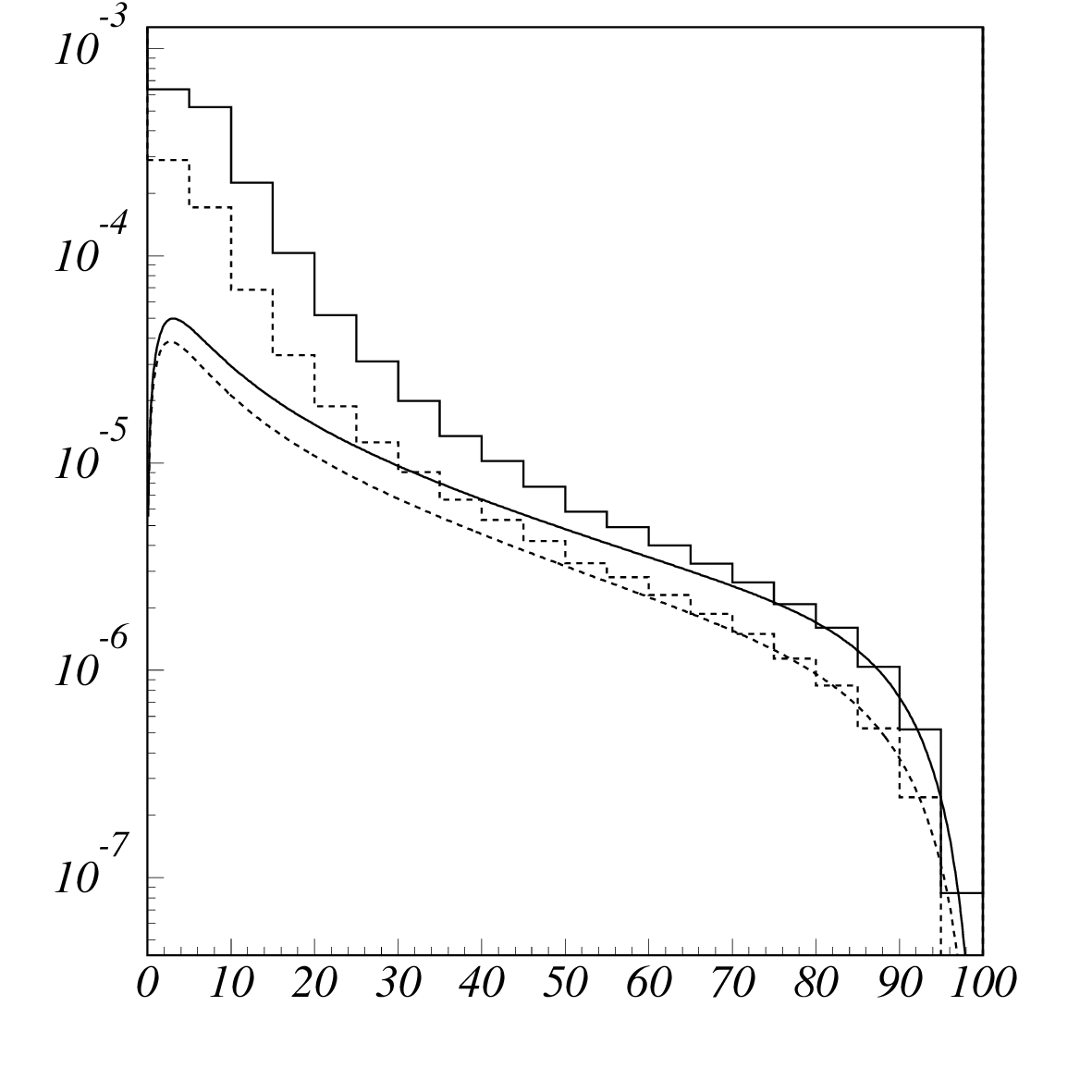}
\end{center}
\vspace*{-13mm}
{\hfill GeV\hspace*{8mm}}
\caption{The differential cross-section for the $B_c^{(*)}$ meson production in
gluon-gluon collisions as calculated in the perturbative QCD over the complete
set of diagrams in the $O(\alpha_s^4)$ order at 200 GeV.
The dashed and solid histograms present the pseudoscalar and vector states, 
respectively, in comparison with the results of fragmentation model
shown by the corresponding smooth curves.}
\label{fig1}
\end{figure}

The $p_T^{min}$ value turns out to be much greater than the $M_{B_c}$ mass, 
so that the dominant contribution into the total cross-section of
gluon-gluon $B_c$-production is given by the diagrams of 
nonfragmentational type, i.e. by the recombination of heavy quarks. 
Furthermore, the convolution of 
the parton cross-section with the gluon distributions inside the initial 
hadrons leads to the suppression of contributions at large transverse 
momenta as well as the subprocesses with large energy in the system of parton 
mass centre, so that the main contribution into the total cross-section of 
hadronic $B_c$-production is given by the region of energies less or
comparable to the $B_c$-meson mass, where the fragmentation model can not be 
applied by its construction. Therefore, one must perform the calculations 
with the account for all contributions in the given order under consideration
in the region near the threshold.

The large numeric value of 
$p_T^{\rm min}$ points out the fact that the basic amount of events of the 
hadronic $B_c^{(*)}$-production does not certainly allow the description in
the framework of the fragmentation model. This conclusion looks more evident, 
if one considers the $B_c$-meson spectrum over the energy.

\begin{figure}[p]
\hspace*{25mm}$\sigma^{-1}_{b\bar b}d\hat\sigma/dz \times 10 ^{-2}$\\
\vspace*{-9mm}
\begin{center}
\hspace*{-10mm}
\epsfxsize=14cm \epsfbox{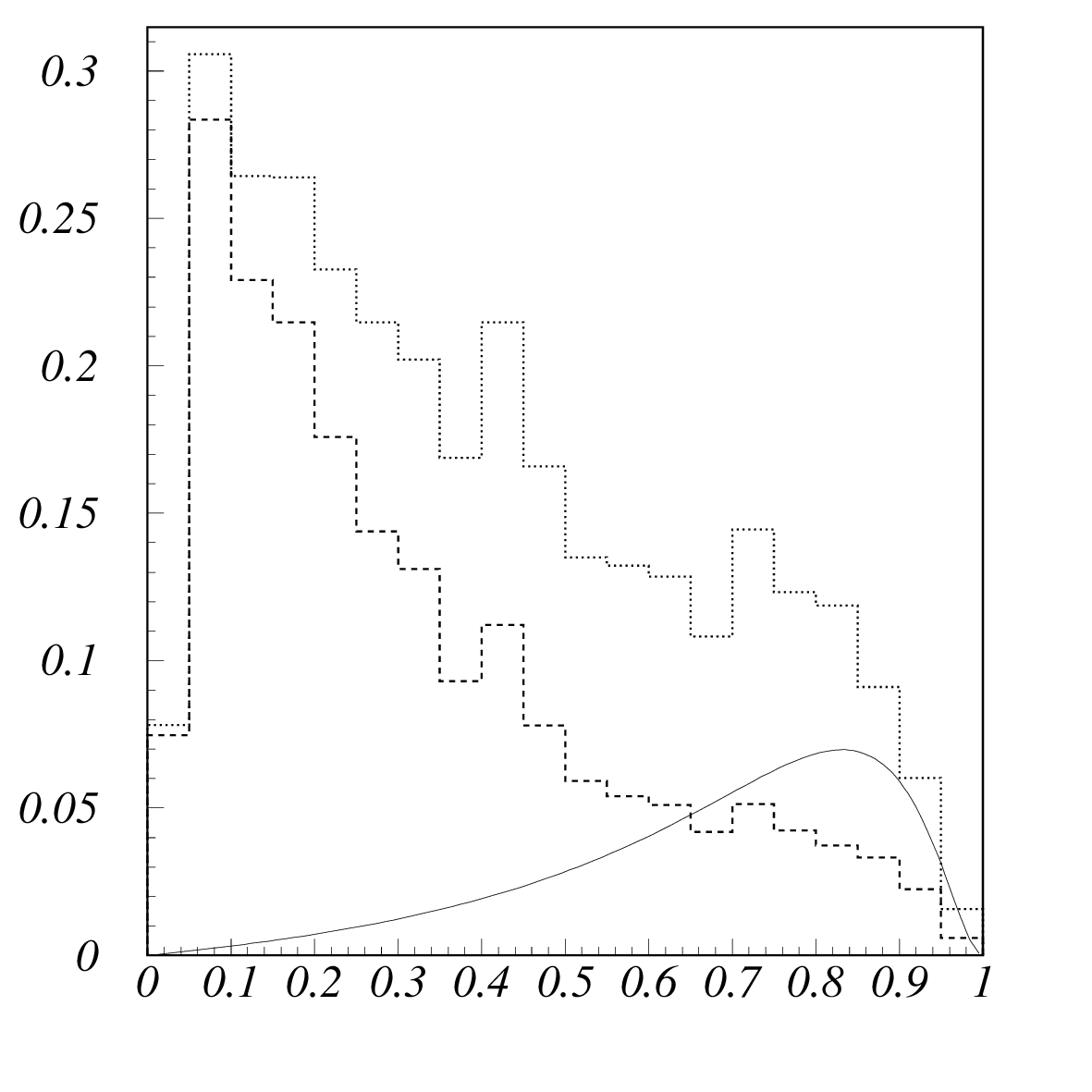}
\end{center}
\vspace*{-13mm}
{\hfill $z$\hspace*{8mm}}
\caption{$d\hat \sigma/dz$ for the $B_c$-mesons at the gluon 
interaction energy of 100 GeV.
The dashed histogram presents the $gg\to B_c +\bar b+ c$ process, the dotted
one is the abelian case. The curve shows the result of the fragmentation model.
The cross-sections are normalized over the cross-section of the $b\bar b$ pair
production.}
\label{pertfig5}
\end{figure}

The basic part of events for the gluon-gluon production of
$B_c$ is accumulated in the region of low $z$ close to 0, where the 
recombination being essentially greater than the fragmentation, dominates.
One can draw the conclusion on the  essential 
destructive interference in the region of $z$ close to 1, for the pseudoscalar
state.

We have in detail considered the contributions of
each diagram in the region of $z\to 1$. In the covariant Feynman gauge the
diagrams of the gluon-gluon production of $Q+\bar Q$ with the subsequent
$Q\to (Q\bar q)$ fragmentation dominate as well as the diagrams,
when the $q\bar q$ pair is produced in the region
of the initial gluon splitting. 
However, the contribution of the latter diagrams
leads to the destructive interference 
with the fragmentation amplitude, and this results in the "reduction"
of the production cross-section in the region of $z$ close to 1.
In the axial gauge with the vector
$n^\mu=p^\mu_{\bar Q}$ this effect of the interference still manifests itself
brighter, since the diagrams like the splitting of gluons 
dominate by several orders of magnitude over the fragmentation, but the
destructive interference results in the cancellation of such extremely
large contributions. This interference is caused by the 
nonabelian nature of QCD, i.e., by the presence of the gluon self-action 
vertices.

To stress the role of the interference diagrams related to the nonabelian 
self-action of gluons, we have considered the process with abelian currents. 
In the abelian case the effect of the 
destructive interference due to the additional contribution of the self-action
of gauge quanta, is absent. So, the agreement between the factorized
model of fragmentation and the exact perturbative calculation is quite good at
$z$ close to 1.

The direct verification of the given mechanism for the $B_c$-meson production
could be the comparison of the $B_c$-meson spectra in two hemispheres in 
the region of the gluon fragmentation and in the photon one, in the 
photonic production of $B_c$ on nucleons.

\begin{figure}[t]
\hspace*{5mm}$d\sigma_{tot}/dp_T$, pb/GeV\\
\vspace*{-9mm}
\begin{center}
\hspace*{-10mm}
\epsfxsize=9cm \epsfbox{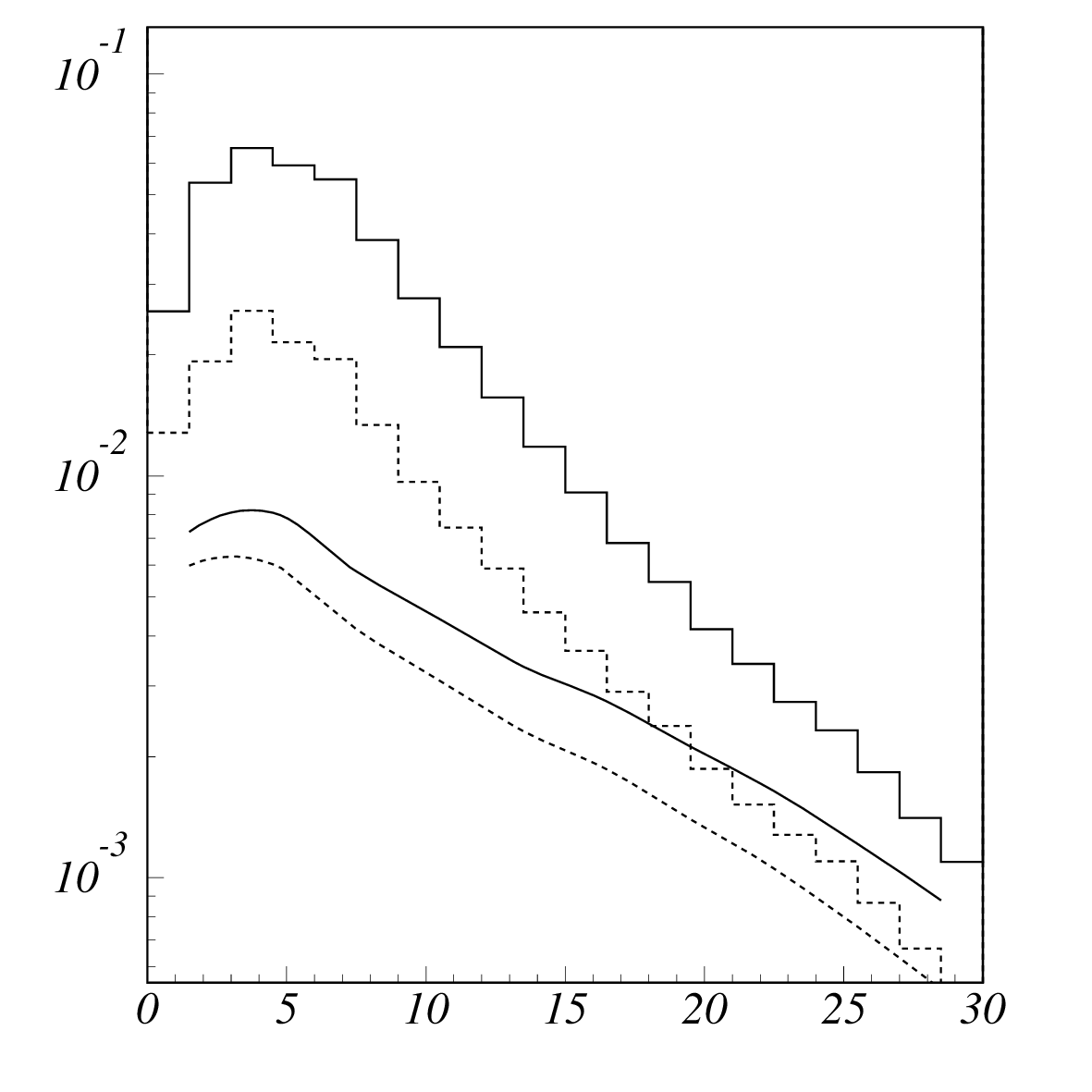}
\end{center}
\vspace*{-13mm}
{\hfill GeV\hspace*{8mm}}
\caption{The differential cross-sections for the $B_c(B_c^*)$-meson
production in $p\bar p$-collisions versus the transverse momentum
at the interaction energy 1.8 TeV with the cut off the gluon-gluon energy
$E>60$ GeV.}
\label{g3fg4a}
\end{figure}
\vspace*{7mm}
At LHC with the luminosity ${\cal L}=10^{34}$ cm$^{-2}$s$^{-1}$ and
$\sqrt{s}=14$ TeV one could expect \cite{bclhc}
$$
4.5\cdot 10^{10}\;\; B_c^+\; \mbox{\sf events per year,}
$$
that leads at the efficiency of the decay reconstruction $\epsilon=0.1$ to
$$
3\cdot 10^4\;\; B_c^+\to \psi\pi^+\; \mbox{\sf events per year,}
$$
with no taking into account some detector-design cuts off angles and momenta.

\newpage
\centerline{\bf\sc  6. Conclusions}

\vspace*{0.4cm}
\begin{itemize}
\item
The family of $(\bar b c)$ mesons contains 16 narrow states, the excited ones
decay into the ground pseudoscalar state due to the radiative cascades.
\item
The mass of ground state is expected to have the value
$$
m_{B_c} =  6.25\pm 0.03\;\; {\rm GeV,}
$$
which is quite close to the value of $B_c\to \psi \pi$ candidates at LEP
$$
[m_{B_c}]_{\rm LEP}  =  6.33\pm 0.05\;\; {\rm GeV.}
$$
\item
The $B_c^+$ meson is the long-lived particle with 
the predicted lifetime equal to
$$
\tau_{B_c}  =  0.55\pm 0.15\;\; {\rm ps,}
$$
which must be compared with the LEP measurements of the candidates
$$
[\tau_{B_c}]_{\rm LEP} =  0.28^{+0.10}_{-0.20}\;\; {\rm ps.}
$$
\item
In $e^+e^-$-annihilation, the fragmentation of $\bar b \to B_c^{(*)+}$
dominates, so that one expects
$$
[f(\bar b \to B_c^+)\cdot {\rm BR}(B_c^+\to \psi \pi^+)]_{\rm TH} =
(0.22\pm 0.09)\cdot 10^{-5},
$$
which is less than the
$B_c^+\to \psi \pi^+$ candidates rate at LEP
$$
[f(\bar b \to B_c^+)\cdot {\rm BR}(B_c^+\to \psi \pi^+)]_{\rm LEP} =
(1.9^{+2.5}_{-1.2}\pm 0.3)\cdot 10^{-5}.
$$
\item
The fragmentation regime works in the hadronic production of
$B_c$ at high transverse momenta $p_T> 35$ GeV only. The recombination
is essential at lower momenta. In the forward production of $B_c$ along the
hadron beam, the destructive interference of fragmentation diagrams with the
gluon-splitting ones takes place.
\item
The most promising production rate of $B_c$ for its observation is
expected in hadron-hadron collisions, especially at LHC and upgraded Tevatron.
\end{itemize}

\end{document}